\documentclass[journal,compsoc]{IEEEtran}
\usepackage{amsmath}
\usepackage{amssymb}
\usepackage{amsthm}
\usepackage{mathrsfs}
\usepackage{dsfont}
\usepackage{stmaryrd} % For \llbracket and \rrbracket
\newtheorem{theorem}{Theorem}
\usepackage{geometry}
\usepackage[utf8]{inputenc}
\usepackage{pgf-pie}
\definecolor{customgray}{RGB}{169, 169, 169}
\definecolor{custommaroon}{RGB}{128, 0, 0}
\definecolor{lighterblue}{rgb}{0.4, 0.6, 0.8}
\definecolor{lighterorange}{rgb}{1, 0.6, 0.4}
\definecolor{lighteryellow}{rgb}{1, 1, 0.6}
\definecolor{lightergray}{rgb}{0.6, 0.6, 0.6}
\definecolor{lightermaroon}{rgb}{0.8, 0.4, 0.4}

\usepackage{dirtytalk}
\usepackage[nottoc]{tocbibind}
\usepackage{graphicx}
\usepackage{xcolor}
\usepackage{listings}
\usepackage{array}
\newcolumntype{L}[1]{>{\raggedright\arraybackslash}p{#1}}
\newcolumntype{R}[1]{>{\raggedleft\arraybackslash}p{#1}}
\newcolumntype{Y}{>{\raggedright\arraybackslash}X}

\usepackage{booktabs}
\usepackage{tabularx} % For tables with adjustable column widths
\usepackage{url}
\usepackage{hyperref}
\usepackage{enumitem} % For itemize in tables
\setlist[itemize]{wide=0pt, leftmargin=*, itemsep=0.5\baselineskip, parsep=0pt, before=\vspace{0.5\baselineskip}, after=\vspace{0.5\baselineskip}}
\usepackage{wrapfig,lipsum,booktabs}
\usepackage{algorithm}
\usepackage{algpseudocode}
\usepackage{tikz}
\usepackage{pgfplots}
\pgfplotsset{compat=1.18}
\usepackage{makecell}
\usepackage{float}
\usepackage{pgf-pie}
\usepackage{subcaption}
\usepackage[numbers]{natbib}
\usepackage{etoolbox}
\usepackage{multirow}

\definecolor{verylightgray}{rgb}{.97,.97,.97}
\definecolor{commentcolor}{rgb}{0.0, 0.6, 0.0}  
\definecolor{keywordcolor}{rgb}{0.0, 0.0, 1.0}  
\definecolor{stringcolor}{rgb}{0.58, 0.0, 0.82}

\lstdefinelanguage{Solidity}{
  keywords={typeof, new, true, false, function, return, null, catch, switch, var, if, in, while, do, else, case, break, contract, public, private, internal, external, payable, address},
  keywordstyle=\color{keywordcolor}\bfseries,
  ndkeywords={class, export, boolean, throw, implements, import, this},
  ndkeywordstyle=\color{gray}\bfseries,
  identifierstyle=\color{black},
  sensitive=false,
  comment=[l]{//},
  morecomment=[s]{/*}{*/},
  commentstyle=\color{commentcolor}\ttfamily,
  stringstyle=\color{stringcolor}\ttfamily,
  morestring=[b]',
  morestring=[b]"
}

\lstdefinelanguage{Java}{
  keywords={abstract, assert, boolean, break, byte, case, catch, char, class, const, continue, default, do, double, else, enum, extends, false, final, finally, float, for, if, implements, import, instanceof, int, interface, long, native, new, null, package, private, protected, public, return, short, static, strictfp, super, switch, synchronized, this, throw, throws, transient, true, try, void, volatile, while},
  keywordstyle=\color{keywordcolor}\bfseries,
  ndkeywords={@Override},
  ndkeywordstyle=\color{gray}\bfseries,
  identifierstyle=\color{black},
  sensitive=true,
  comment=[l]{//},
  morecomment=[s]{/*}{*/},
  commentstyle=\color{commentcolor}\ttfamily,
  stringstyle=\color{stringcolor}\ttfamily,
  morestring=[b]',
  morestring=[b]"
}

\AtBeginEnvironment{thebibliography}{\fontsize{7.5}{9.5}\selectfont}

\title{Strengthening Solidity Invariant Generation: From Post- to Pre-Deployment}

\author{
  \IEEEauthorblockN{
    Kartik Kaushik,
    Raju Halder,
    Samrat Mondal
  } \\
\IEEEauthorblockA{\textit{Indian Institute of Technology Patna, Patna, India}\\
\textit{Email: \{kartik\_2221cs32, halder, samrat\}@iitp.ac.in}}}
\date{Dec 2023}

\begin{document}

\maketitle

\begin{abstract}
Invariants are essential for ensuring the security and correctness of Solidity smart contracts, particularly in the context of blockchain's immutability and decentralized execution. This paper introduces \textsf{InvSol}, a novel framework for pre-deployment invariant generation tailored specifically for Solidity smart contracts. Unlike existing solutions, namely \emph{InvCon}, \emph{InvCon+}, and \emph{Trace2Inv}, that rely on post-deployment transaction histories on Ethereum mainnet, \textsf{InvSol} identifies invariants before deployment and offers comprehensive coverage of Solidity language constructs, including loops. Additionally, \textsf{InvSol} incorporates custom templates to effectively prevent critical issues such as reentrancy, out-of-gas errors, and exceptions during invariant generation. We rigorously evaluate \textsf{InvSol} using a benchmark set of smart contracts and compare its performance with state-of-the-art solutions. Our findings reveal that \textsf{InvSol} significantly outperforms these tools, demonstrating its effectiveness in handling new contracts with limited transaction histories. Notably, \textsf{InvSol} achieves a 15\% improvement in identifying common vulnerabilities compared to \emph{InvCon+} and is able to address vertain crucial vulnerabilities using specific invariant templates, better than \emph{Trace2Inv}.

\end{abstract}

\section{Introduction}

Solidity, the smart contract language for Ethereum, has emerged as a leading choice among blockchain developers. Given the critical nature of decentralized applications and blockchain immutability, correctness guarantees of Solidity codes before their deployment is paramount. In this context, generating invariants, which hold true in every execution of the code, plays a pivotal role in addressing crucial software engineering tasks including code comprehension, testing/verification, and error/vulnerabilities diagnosis.

Despite such critical importance of Solidity invariant generation, the current literature presents only three notable proposals, namely \emph{InvCon} \cite{liu2022invcon}, \emph{InvCon+} \cite{liu2024automated} and \emph{Trace2Inv} \cite{chen2024demystifying}. \emph{InvCon} performs dynamic invariant detection for post-deployed smart contracts using Daikon by feeding data traces generated from the transactions history of the smart contract on Ethereum mainnet. Subsequently, \emph{InvCon+} advances upon the previous work, \emph{InvCon}, by combining both dynamic and static techniques. In particular, the invariant generated dynamically using transaction data of the smart contracts are statically verified using Houdini-like algorithm via VeriSol, with redundant invariants pruned using the Z3 Solver. \emph{Trace2Inv} generates customized invariants for smart contracts by analyzing historical transaction data to detect and mitigate specific vulnerabilities.

We examine all these approaches, closely and observe the following limitations:
\begin{itemize}
    \item \emph{Dependence on Post-Deployment Transaction Histories}: Existing approaches rely on the post-deployment transaction history of smart contracts to generate their invariant. This poses significant limitations to their applicability, especially concerning static code verification aimed at detecting and fixing bugs in smart contracts before deployment. In fact, post-deployment error resolution in smart contracts is hindered by the blockchain ledger's immutability, adding complexity to the process.
    
    \item \emph{Performance Issues for Newly Deployed Contracts}: The absence of transaction history data, in case of newly deployed or less frequently used smart contracts, often leads to the creation of inefficient invariants that may overlook specific behaviors or edge cases. This may render the smart contracts prone to errors and security breaches \cite{nikolic2018finding}.
    
    \item \emph{Inadequate support for loop invariant detection}: None of these existing proposals adequately detect loop invariants. Loops are important in a programming language as they handle repetitive tasks. In a financially critical and operationally crucial programming language like Solidity, behaviour of loops cannot be ignored. For an instance, in case of withdrawal of balance from accounts, a wrongly implemented loop might deplete the entire treasury, causing irreversible circumstances. For example, consider the \texttt{distribute} function of the smart contract shown in Figure \ref{fig:distribute_function} that  iterates over the list of payees, and transfers a percentage of the contract's total balance to each based on their predefined shares. This percentage is represented by the entry in the \texttt{shares} mapping. A crucial property which needs to assert here is that the sum of all shares should not exceed 100\%. Otherwise, the contract would end up paying more than the available balance and ambiguities might occur. But there is no check to ensure so. In this regard, a crucial loop invariant could be:
    \[
    \text{@loop\_invariant: } \sum_{i=0}^{n-1} \text{shares}[i] \leq 100;
    \]
    Currently available tools fail to detect such crucial loop invariants. Therefore, devising of a methodology to detect these loop invariants can greatly enhance the safety and security of smart contracts.
    %
    %distributes funds among different stakeholders. The contract operates on a number of payess and their respective shares in the total contract balance.
    %\todo{need to make the following sentences more attractive: It is notable that loops are pivotal in programming languages and loop invariants are of high relevance \cite{bradley2007calculus}, \cite{hoare1969axiomatic} especially in financially sensitive technologies like blockchain technology. Therefore there is a predominant need of such an invariant generator tool which can detect loop invariants also, in Solidity smart contracts. }
\end{itemize}

\begin{figure}[t]
    \centering\scriptsize
    \begin{lstlisting}[language=Solidity]
    function distribute() public {
        for (uint i = 0; i < payees.length; i++) {
        address payee = payees[i];
        uint256 payment = address(this).balance * shares[payee] / 100;
        payable(payee).transfer(payment);
        }
    }
    \end{lstlisting}
    \caption{A code snippet for distribute function}
    \label{fig:distribute_function}
    \end{figure}
    
In this paper, we present \textsf{InvSol}, a state-of-the-art framework for the synthesis and meticulous validation of invariants from Solidity smart contracts prior to their deployment. \textsf{InvSol} achieves this by introducing a Solidity-to-Java language semantic transcompilation mechanism, which not only maintains semantic allegiance of original Solidity script but also allows an automated code annotation using Java Modeling Language (JML). This ensures enrichment of the Solidity codes with exact formal specifications to capture both behavioural and security related aspects of the smart contracts. At the core, a dynamic invariant discovery module, along with an appropriate test crafter, carries out invariant generation with the help of standard invariant schemas enhanced with customized invariant templates to capture blockchain's distinguished operational and gas related mechanisms. Furthermore, an adaptive and iterative feedback mechanism is equipped along with static analysis which ensures correctness of the generated invariants.  

We thoroughly evaluate \textsf{InvSol} for its capability to generate insightful invariants on a set of real world smart contracts. We employ a dual strategy for this evaluation. First, we experiment \textsf{InvSol} on 493 Solidity smart contracts of varied types and complexities collected from \emph{Google BigQuery} \cite{google_bigquery} platform, in order to assess the correctness of invariants and their coverage across diverse language constructs. Next, we run \textsf{InvSol} on \emph{SmartBugs Curated} \cite{smartbugs_dataset} dataset which comprises 143 vulnerable smart contracts, aiming at determining whether \textsf{InvSol's} invariants effectively address these vulnerabilities and assist in mitigating potential security issues. Finally we evaluate \textsf{InvSol} on benchmark dataset used by Chen et al. for evaluating \emph{Trace2Inv}, which comprised of 42 smart contracts.

In summary, our primary contributions in this paper are:
\begin{itemize}
    \item We introduce \textsf{InvSol}, a novel framework to generate invariants in Solidity smart contracts prior to their deployment. Unlike the existing approaches, \textsf{InvSol} eradicates the dependency on transaction history of Solidity smart contracts for invariant generation.
    \item We enhance \textsf{InvSol's} ability, enabling it to maximize the code coverage, specially rooting for blockchain specific characteristics such as gas costs, reentrancy, and exception handling. It is also able to generate loop invariants, which no other current tools are able to detect.
    \item We rigorously evaluate \textsf{InvSol's} ability to generate meaningful invariants across 678 real-world smart contracts, sourced from \emph{Google BigQuery}, \emph{SmartBugs} and \cite{chen2024demystifying}. The performance assessment reveals that \textsf{InvSol} performs better than the existing invariant generation tools and effectively addresses loop invariant. Alongside, it achieves on an average 15\% better in case of addressing vulnerabilties-related invariants.
\end{itemize}

The rest of the paper is organized as follows: Section \ref{label:Hoare-logic} discusses the relevance of invariants in ensuring program correctness. Section \ref{label:invariant-synthesis} describes the architecture of our proposed invariant generation framework and the operational details of its various modules. Section \ref{label:experimental-evaluation} presents experimental results and performance comparison against state-of-the-art approaches. Section \ref{label:related-works} provides an overview of the related work. Finally, Section \ref{label:conclusion} concludes the work.

\section{Hoare Logic and Smart Contract Correctness}
\label{label:Hoare-logic}
Floyd-Hoare logic \cite{hoare1969axiomatic} is a formal system for reasoning of the correctness of computer programs. It is an effective way to ensure the reliability of smart contracts. This employs preconditions and postconditions denoted as
\begin{center}
\{P\}Prog\{Q\}
\end{center}
where \{P\} is the precondition, Prog is the program, and \{Q\} is the postcondition. The Floyd-Hoare logic asserts that if the execution of Prog from the states satisfying P terminates then the resultant states satisfy Q. 

\begin{figure}[t]
\centering
\begin{lstlisting}[language=Solidity]
contract AssetTransfer {
    address public owner;

    constructor() {
        owner = msg.sender; // { true }
    }

    function transferOwnership(address newOwner) public {
        // { msg.sender == owner }
        require(msg.sender == owner, "Only the owner can transfer ownership.");
        owner = newOwner;
        // { owner == newOwner }
    }
\end{lstlisting}
\caption{Solidity code for transferring asset ownership, ensuring that only the owner can initiate the transfer.}
\label{fig:asset_transfer}
\end{figure}

Consider the code snippet in Figure \ref{fig:asset_transfer} which comprises a simple Solidity function \texttt{transferOwnership} that transfers ownership of an asset. This function is annotated with the preconditions \(\{ msg.sender == owner \}\) and the postconditions \(\{ owner == newOwner \}\), demonstrating a practical application of Floyd-Hoare logic. These annotations represent formal specifications stating that if the owner executes this function (pre-condition), then on termination, the ownership is transferred to the newOwner (post-condition). This is an instance showing that if the precondition is met, upon execution and termination, the post condition is eventually followed. This means that the program execution adheres correctly to the specification. The need of program invariants stems in the context of Floyd-Hoare logic to help in ensuring correct program behaviour and consistent state transfers. Such annotated formal specifications act as vital references for automated verification tools, enabling them to confirm the correctness of the contract's operations.

\subsection{Contract Vulnerabilities and Invariants}
\label{label:contract_vulnerabilties}
Solidity smart contracts, which play a vital role in the implementation of decentralized applications, often get subjected to inconsistencies or risks like reentrancy, overflow/underflow, etc. Therefore, extensive formal methods, especially the one which generate invariant properties, become crucial to ensure safety and security. 

A \textbf{Safety Property} in the context of smart contracts ensures that the state of the system never reaches a ``bad'' or undesirable state throughout the execution process. This property prevents the system from going to states which could lead to attacks or security breaches, thus maintaining security and consistency in the system. Formally, we define a safety property as:
\[
\text{Safety}(\Delta) := \forall \rho, \rho'\in \Sigma, (\rho \xrightarrow{\vec{a}} \rho')\in \Delta \Rightarrow \neg P(\rho')
\]
Where, \( \Delta \) represents the set of all possible state transitions within the contract, \( \xrightarrow{\vec{a}} \) denotes state transition function that maps a state \( \rho \) to a new state \( \rho' \) based on the vector of actions or inputs \( \vec{a} \). \( P \) represents a predicate that defines an undesirable state. The definition asserts that for any state transition within the contract, it is impossible to transit into a state \( \rho' \) where \( P \) holds true.

Invariants ensure this safety by maintaining constant conditions:
\[
\forall \rho, \rho'\in \Sigma, \: \rho \xrightarrow{\vec{a}} \rho' \Rightarrow (I(\rho) \land I(\rho'))
\]
Here, \( I \) represents the invariant conditions, which should remain true throughout the program execution. These invariants ensure that if they hold in an initial state \( \rho \), they must also hold in the subsequent state \( \rho' \), and hence prevent the contract from entering any undesirable state.

Practically, these invariants play a crucial role in Solidity smart contracts by safeguarding against overflow/underflow in transaction values, upholding the logic of token transfers, securing access control mechanisms, and performing other similar activities \cite{so2020verismart}. By preventing going in any such state where the invariant might prove false, many such unsafe states are blocked which might lead to unwanted functionalities.

\textbf{Loop Invariants} are a particular type of invariants which are used to analyze and secure loop behaviours in a programming language. A loop invariant must hold before and after each iteration of the loop, and in this way help in securing overall correctness of programs having loops. Formally, a loop invariant can be introduced as follows: Given a loop \( l \) and a set of initial states $\mathcal{S}_{0}$, the state transition semantics of \( l \) can be defined as a fix-point solution of $F$ where
\[
F = \lambda_{\psi} \cdot \mathcal{S}_{0} \bigcup \{ \rho' \mid \rho \in \psi, \rho \rightarrow \rho' \}
\]
We can define \( I_{l} \) as a loop invariant if:
\[
\forall i \geq 0 : I_{l}(F_{i}) \text{ holds where } F_{0} = \mathcal{S}_{0}.
\]

\section{Invariant Synthesis and Execution}
\label{label:invariant-synthesis}
\subsection{Overview}

\begin{figure}[t]
\centering
\includegraphics[width=8cm]{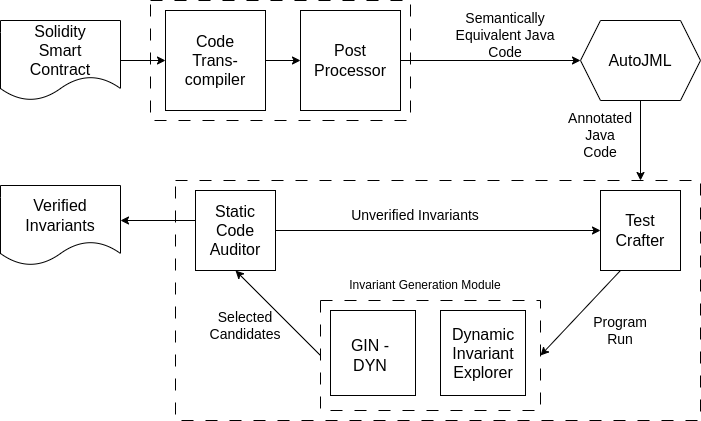}
\caption{\textsf{InvSol} System Architecture}
\label{fig:architecture_solicon}
\end{figure}

Figure \ref{fig:architecture_solicon} shows the architecture of \textsf{InvSol}. It comprises several interconnected modules. The \textit{Code Transcompiler} module transcompiles Solidity code into Java equivalent while maintaining the original semantics. It is followed by the \textit{Post Processor} module, which augments the generated code with a number of custom classes simulating Solidity-specific features and blockchain environment. The \textit{AutoJML} eventually enriches this Java code with JML annotations to facilitate invariant detection. To generate a rich set of test cases ensuring a comprehensive coverage of the code, the \textit{Test Crafter} module adopts \emph{Evosuite} \cite{evosuite} test case generator after tailoring it with custom solidity-specific features so as to check gas consumption, reentrancy and other exceptions. Inside the \textit{Invariant Generator Module}, the \textit{Dynamic Invariant Explorer} then processes the prepared code to generate potential invariants, while the \textit{GIN-DYN} focuses specifically on identifying loop invariants. Finally, the \textit{Static Code Auditor} statically verifies these candidate invariants, confirming their correctness. When the \textit{Static Code Auditor} is unable to verify some candidates, the \textit{Test Crafter} module is initiated again to generate a set of new tests to disprove these candidates.

\begin{algorithm}
\caption{Invariant Generation Process}
\label{alg:intraprocess}
\begin{algorithmic}[1]
    \Procedure{\textit{GenInvariants}}{\texttt{SolidityCode}}
        \State $JC \gets \text{\texttt{Translate}}(\texttt{SolidityCode})$
        \State $AJC \gets \text{\textit{AutoJML}}(JC)$
        \State $TS \gets \text{\texttt{GenTests}}(AJC)$
        \State $CandInv \gets \text{\texttt{ExploreInvs}}(TS)$
        \State $VerInv \gets \text{\texttt{VerifyInvs}}(CandInv)$
        \While{not \text{\texttt{IsFixed}}(\texttt{VerInv})}
            \State $AddTests \gets \text{\texttt{RefineTests}}(\texttt{VerInv})$
            \State $TS \gets TS \cup AddTests$
            \State $CandInv \gets \text{\texttt{ExploreInvs}}(TS)$
            \State $VerInv \gets \text{\texttt{VerifyInvs}}(CandInv)$
        \EndWhile
        \State \textbf{return} $VerInv$
    \EndProcedure
\end{algorithmic}
\end{algorithm}

We summarize the entire invariant generation procedure in Algorithm \ref{alg:intraprocess}. It begins by transcompiling the Solidity code into Java (\emph{JC}) equivalent, using the \texttt{Translate} function, which preserves semantic integrity and ensures that the generated Java code accurately reflects the behaviors and logic defined in the original Solidity script. Following this, the Java code (\emph{JC}) is automatically annotated with Java Modeling Language (JML) annotations through the \textit{AutoJML} function, enriching the code with formal specifications that aid in subsequent invariant analysis and testing. Test cases (\emph{TS}) are then generated for the annotated Java code (\emph{AJC}) by the \texttt{GenTests} function, aimed at maximizing coverage and ensuring that the annotations validate against the code's runtime behavior and state changes. The algorithm proceeds to explore and verify potential invariants from these test cases using the \texttt{ExploreInvs} and \texttt{VerifyInvs} functions, confirming the correctness of candidate invariants (\texttt{CandInv}) that describe the program's behavior across all tested scenarios.

In the loop, the algorithm refines these test cases and revisits invariant exploration and verification until a fixed point (\texttt{IsFixed}) is achieved where no new invariant is discovered. The verified invariants (\emph{VerInv}) are then finalized as accurate behavioral descriptors of the original Solidity codes, obtained though a systematic transcompilation, annotation, and verification process in the Java environment.

Now we discuss each of these modules in a greater detail.

\subsection{Code Transcompiler}

This module is responsible to transcompile Solidity code into semantically equivalent Java code, and hence serves as the foundation for invariant detection process.

The process of begins with the \textit{Solidity Compiler} (\texttt{solc}) generating an Abstract Syntax Tree (AST) from the Solidity source code. This AST is refined to emphasize essential nodes that significantly influence the transcompilation process.

Initially, \texttt{SourceUnit} nodes are identified as the root, which contain all contract information, while non-essential nodes such as \texttt{ImportDirective} and \texttt{PragmaDirective} are discarded to streamline the AST. This ensures that only nodes critical for accurate code translation, such as \texttt{ContractDefinition}, are retained.

Subsequently, the retained AST nodes are mapped to Java constructs. This process employs a depth-first traversal of the AST, ensuring that each node is appropriately transformed based on its type and hierarchy within the AST. For instance, \texttt{FunctionDefinition} nodes are converted into Java methods, incorporating their associated parameters and modifiers to maintain behavioral fidelity.

The transcompilation of state variables and functions is handled by mapping Solidity types to their closest Java equivalent types or by applying custom classes wherever necessary so as to preserve integrity and operational behavior. The aim is to ensure that the resultant Java code faithfully replicates the logic and structure of the original Solidity contracts, thereby enabling effective invariant analysis. Table \ref{tab:mapping_solidity_java} shows mapping of general constructs from Solidity to Java. This module currently supports Solidity compiler versions ranging between 0.4.22 and 0.6.0.

\begin{table*}[t]
\centering\tiny
\caption{Mapping of general constructs between Solidity and Java}
\label{tab:mapping_solidity_java}
%\resizebox{\textwidth}{!}{%
\begin{tabular}{@{}p{0.15\textwidth}p{0.4\textwidth}p{0.45\textwidth}@{}}
\toprule
\textbf{Solidity AST Node} & \textbf{Solidity Construct} & \textbf{Equivalent Java Construct} \\\midrule
SourceUnit & \texttt{pragma solidity \textasciicircum{}$\ll$version$\gg$;  $\dots$ } & Treated as the entry point of the Java code\\
ContractDefinition & \texttt{contract $\ll$contract-name$\gg$ \{ \dots \}} & \texttt{class $\ll$contract-name$\gg$ \{ \dots \}} \\
ElementaryTypeName & \texttt{$\ll$Solidity-datatype$\gg$} & \texttt{$\ll$Equivalent Java-datatype$\gg$} \\
State-VariableDeclaration & \texttt{{}$\ll$Solidity-datatype$\gg$ {}$\ll$access-modifier$\gg$ {}$\ll$variable-name$\gg$} & \texttt{{}$\ll$access-modifier$\gg$ {}$\ll$Equivalent Java-datatype$\gg$ {}$\ll$variable-name$\gg$} \\
Parameter-VariableDeclaration & \texttt{$\ll$Solidity-datatype$\gg$ \texttt{$\ll$data-location$\gg$} $\ll$variable-name$\gg$} & \texttt{{}$\ll$Equivalent Java-datatype$\gg$ \texttt{$\ll$custom data-location class$\gg$} $\ll$variable-name$\gg$} \\
Constructor & \texttt{constructor($\ll$parameters$\gg$) \{ \dots \}} & \texttt{$\ll$contract-name$\gg$($\ll$Mapped Parameters$\gg$) \{ \dots \}} \\

FunctionDefinition & \texttt{function $\ll$function-name$\gg$($\ll$parameters$\gg$) $\ll$access-modifier$\gg$ [pure|view|payable] [returns ($\ll$return types$\gg$)] \{ \dots \}} & \texttt{$\ll$access-modifier$\gg$ $\ll$return type$\gg$ $\ll$function-name$\gg$($\ll$parameters$\gg$) \{ \dots \}} \\
FunctionCall & \texttt{$\ll$function-name$\gg$({}$\ll$parameters$\gg$)} & \texttt{$\ll$function-name$\gg$($\ll$parameters$\gg$)} \\
Assignment & \texttt{$\ll$variable-name$\gg$ = $\ll$expression$\gg$} & \texttt{$\ll$variable-name$\gg$ = $\ll$Equivalent expression$\gg$} \\
Return & \texttt{return {}$\ll$variable-name$\gg$} & \texttt{return $\ll$variable-name$\gg$} \\
ForStatement & \texttt{for ($\ll$initialization$\gg$; $\ll$condition$\gg$; $\ll$iteration$\gg$) \{ \dots \}} & \texttt{for ($\ll$ Equivalent initialization$\gg$; $\ll$Equivalent condition$\gg$; $\ll$Equivalent iteration$\gg$) \{ \dots \}}\\
While & \texttt{while ($\ll$condition$\gg$) \{ \dots \}} & \texttt{while ($\ll$Equivalent condition$\gg$) \{ \dots \}} \\
If-Else & \texttt{if ($\ll$condition$\gg$) \{ \dots \} else \{ \dots \}} & \texttt{if ($\ll$Equivalent condition$\gg$) \{ \dots \} else \{ \dots \}} \\
Inheritance & \texttt{contract $\ll$derived-name$\gg$ is $\ll$base-name$\gg$, $\ll$interface-names$\gg$ \{ \dots \}} & \texttt{class $\ll$derived-name$\gg$ extends $\ll$base-name$\gg$ implements $\ll$interface-names$\gg$ \{ \dots \}} \\
\bottomrule
\end{tabular}
%}
\end{table*}

\subsection{Post Processor}

\begin{figure}[t]
\centering
\includegraphics[width=8cm]{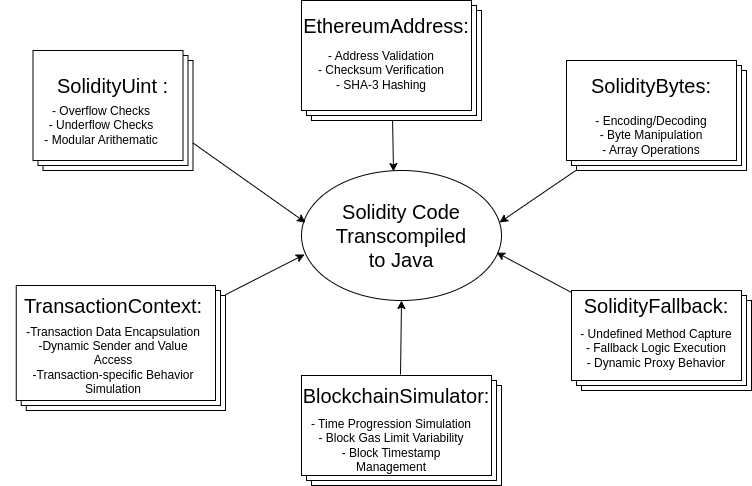}
\caption{Overview of custom classes in Post Processor}
\label{fig:post_proc_custom}
\end{figure}

This phase refines the initially translated Java code to accurately simulate Solidity's unique blockchain features and semantics. This critical phase ensures the final Java code is not only syntactically correct but also preserves the operational integrity of the original Solidity code.

Initially, Java constructs generated from the AST are woven into coherent code structures. This includes special handling for blockchain-specific elements like \texttt{msg.sender} and \texttt{msg.value}, which are crucial for interactions within the Ethereum network. These elements are integrated using the custom \textit{TransactionContext} class, which replicates the dynamic nature of blockchain transactions. The simulation of blockchain environment characteristics is meticulously handled via a custom \textit{BlockchainSimulator} class, depicted in Listing \ref{lst:blockchainsimulator}. This class enhances the simulation by adjusting blockchain time and gas limits dynamically, mirroring the Ethereum model. 

Figure \ref{fig:post_proc_custom} gives a glimpse of different custom classes added by this phase and the functionalities of each custom class. 

\begin{lstlisting}[language=Java, caption={An excerpt of custom BlockchainSimulator class.}, label={lst:blockchainsimulator}]
public class BlockchainSimulator {
    private Instant currentTimestamp;
    private long currentGasLimit;
    private final long initialGasLimit;
    private final Random random;
    public BlockchainSimulator(long initialGasLimit) {
        this.initialGasLimit = initialGasLimit;
        this.currentGasLimit = initialGasLimit;
        this.currentTimestamp = Instant.now();
        this.random = new Random();
    }
    public void mineNewBlock() {
        this.currentTimestamp = this.currentTimestamp.plusSeconds(15); 
        long variance = (long) (this.initialGasLimit * 0.1);
        this.currentGasLimit = initialGasLimit + (long) (variance * (2 * random.nextDouble() - 1));
    }
    public Instant getCurrentTimestamp() {
        return this.currentTimestamp;
    }
    public long getCurrentGasLimit() {
        return this.currentGasLimit;
    } // rest of the code    
}
\end{lstlisting}

\begin{figure*}[t] % This ensures the figure spans both columns and is placed at the top of the page
    \centering
    \begin{subfigure}[b]{0.5\textwidth}
        \centering
        \scriptsize % Smaller font size for fitting content
        \[
        \begin{aligned}
            T ::= &\text{address} \mid \text{uint} \mid \text{int} \mid \text{string} \mid \text{bytes} \mid \\
             & \text{byte} \mid \text{bool} \mid \text{array} \mid \text{mapping} \mid \text{struct} \{ \overrightarrow{T} \} \\
            V ::= &\text{identifier} \mid V[V] \mid V.\text{identifier} \\
            F ::= &\text{func}(\overrightarrow{V}) \{ \overrightarrow{S} \} \\
            S ::= & V := E \mid \text{if} (E) \{ \overrightarrow{S} \} \text{ else } \{ \overrightarrow{S} \}  \\
              & \mid \text{while} (E) \{ \overrightarrow{S} \} \mid \text{call}(\overrightarrow{E}) \mid \text{return} E \\
              & \mid\text{require}(E) \mid \text{assert}(E)\\ %\mid \text{revert} \mid \text{init} \\
            E ::= & \text{const} \mid E[E] \mid E.V \mid E \diamond E \mid \text{unary-op}~E \\
         \text{Contract} \ni c ::= &\langle \vec{v}_s, \texttt{initS}, \text{MS}\rangle
        \end{aligned}
        \]
        \caption{Solidity Abstract Syntax}
        \label{fig:solidity-grammar}
    \end{subfigure}%
    \begin{subfigure}[b]{0.5\textwidth}
        \centering
        \scriptsize % Smaller font size for fitting content
        \[
        \begin{aligned}
            T ::=& \text{int} \mid \text{double} \mid \text{boolean} \mid \text{char} \mid \text{String} \mid \text{Class} \mid \text{Array} \\
            V ::=& \text{identifier} \mid V[V] \mid V.\text{identifier} \\
            M ::=& \text{method}(\overrightarrow{V}) \{ \overrightarrow{S} \} \\
            S ::= &V := E \mid \text{if} (E) \{ \overrightarrow{S} \} \text{ else } \{ \overrightarrow{S} \} \mid \text{while} (E) \{ \overrightarrow{S} \} \\
             &\mid \text{System.out.println}(E) \mid \text{return} E \mid \text{try} \{ \overrightarrow{S} \} \text{ catch } (E) \{ \overrightarrow{S} \} \\
            E ::=& \text{const} \mid E[E] \mid E.V \mid E \diamond E \mid \text{new} T(\overrightarrow{E}) \mid \text{unary-op}~E \\
    \text{Class} \ni c ::= &\langle \text{F}, \texttt{initJ}, \text{MJ} \rangle
        \end{aligned}
        \]
        \caption{Java Abstract Syntax}
        \label{fig:java-grammar}
    \end{subfigure}
    \caption{Comparative core grammars of Solidity and Java programming languages}
    \label{fig:grammars}
\end{figure*}

\subsubsection{Transcompilation Formalism}
Figure \ref{fig:grammars} depicts the abstract syntax of Solidity and Java. A smart contract $c$ is defined as a triplet $\langle \vec{v}_s, \texttt{initS}, \text{MS}\rangle$, where \(\vec{v}_s\) represents the set of \textit{variables} that store the contract's persistent state, \(\texttt{initS}\) denotes the contract constructor, and \(\text{MS}\) is the set of \textit{functions} that are responsible for executing behaviors and altering the state of contract. Let $\text{Var}$, $\text{Val}$, $\text{MLoc}$, and $\text{SLoc}$ represent sets of variables, the domain of values, local memory locations, and blockchain storage locations, respectively. We define the notion of Solidity states as follows:
\begin{itemize}
    \item $\text{LEnv}:~\text{Var} \rightarrow \text{MLoc}$ - A mapping from variables to local memory locations.
    \item $\text{LStore}:~\text{MLoc} \rightarrow \text{Val}$ - A mapping from local memory locations to values.
    \item $\text{BEnv}:~\text{Var} \rightarrow \text{SLoc}$ - A mapping from variables to blockchain storage locations.
    \item $\text{BStore}:~\text{SLoc} \rightarrow \text{Val}$ - A mapping from blockchain storage locations to values.
    \item $\Sigma:~\text{LEnv} \times \text{LStore} \times \text{BEnv} \times \text{BStore}$ - A composite state, denoted by $\rho = \langle le, ls, be, bs \rangle$.
\end{itemize}

In Java, a class $c$ is defined as a triplet $\langle \text{F}, \texttt{initJ}, \text{MJ} \rangle$, where \(\text{F}\) represents the set of class \textit{Fields} defining its attributes, \(\texttt{initJ}\) denotes the \textit{Initializer} or \textit{Constructor} responsible for setting up the class instance, and \(MJ\) is the set of \textit{Methods}. Let $\text{Var}$ be the set of variable identifiers, $\text{Val}$ be the domain of values (including primitives and references), and $\text{Loc}$ be the set of memory locations. The Java environments and states are defined below:
\begin{itemize}
    \item $\text{Env}:~\text{Var} \rightarrow \text{Loc}$ - A mapping from variables to memory locations.
    \item $\text{Store}:~\text{Loc} \rightarrow \text{Val}_{\bot}$ - A mapping from memory locations to values, where $\text{Val}_{\bot}=\text{Val}\cup \bot$.
    \item $\Sigma:~\text{Env} \times \text{Store}$ - A composite state of the program, denoted by $\rho = \langle e, s \rangle$.
\end{itemize}

First, let us formalize the state translation semantics. Let $\Sigma_{\text{sol}}$ and $\Sigma_{\text{java}}$ be the set of states corresponding to Solidity and Java, respectively. Given a Solidity state \( \rho_{\text{sol}} = \langle le, ls, be, bs \rangle \in \Sigma_{Sol}\), and an initial java state $\langle e_0, s_0\rangle\in\Sigma_{\text{java}}$,  we define a state translation function $\mathbb{T}_{\rho}:~\Sigma_{\text{sol}}\rightarrow \Sigma_{\text{java}}$, as follows:
\[
\mathbb{T}_{\rho} \left\llbracket \rho_{\text{sol}} \right\rrbracket = \\
\mathbb{T}_{\rho} \left\llbracket \langle le, ls, be, bs \rangle \right\rrbracket = \\
\langle e, s \rangle = \\
\rho_{\text{java}} \in \Sigma_{\text{java}}
\]
such that 
\begin{align*}
    e =&  e_0[ \text{st} \mapsto a, \vec{v}_{\text{l}} \mapsto \vec{a}_{\text{l}}]\\
    s =& s_0 [ a.\vec{F} \mapsto bs(be(\vec{v}_{\text{s}})), \vec{a}_{\text{l}} \mapsto \text{ls}(\text{le}(\vec{v}_{\text{l}}))]
\end{align*}
where
\begin{itemize}
    \item $\vec{v}_{\text{s}}$ and $\vec{v}_{\text{l}}$ are the list of storage variables and local variables in Solidity.
    \item $c = \langle F, \text{init}, M \rangle$ is a custom \textit{TransactionContext} class designed to simulate the blockchain environment within Java, and $a$ is the address where the instance of the class $c$ is stored.
    \item $\text{st}$ is the identifier representing the simulated blockchain storage in Java. 
    \item $\vec{a}_{\text{l}}$ refers to the list of memory addresses that correspond to the local variables in the Solidity environment.

\end{itemize}

Observe that this state translation function simulates the blockchain storage environment and store $\langle be, bs\rangle$ into Java environment and store $\langle e, s\rangle$ by creating an object of the custom class \textit{TransactionContext} and by mapping the storage variables $\vec{v}_{\text{s}}$ to the fields $\vec{F}$ of the object, while keeping the local variables $\vec{v}_{\text{l}}$ intact.

Now, let us move towards language translation semantics. Let $\text{Stmts}_{\text{sol}}$ and $\text{Stmts}_{\text{java}}$ be the set of statements in Solidity and Java languages, respectively. We define a language translation function $\mathbb{T}: \text{Stmts}_{\text{sol}} \rightarrow \text{Stmts}_{\text{java}}$, which translates Solidity constructs into semantically equivalent Java constructs. Let us formalize the translation using a Solidity assignment statement $X := \text{exp}$, where $X$ is a storage variable and $\text{exp}$ is an arithmetic expression involving both local and storage variables. The translation semantics of the above-mentioned assignment statement is defined as:
\[
\mathbb{T} \llbracket X := \text{exp} \rrbracket = (\text{st}.\vec{F}(X) := \text{exp}')
\]
Where, $\vec{F}(\text{id})$ denotes the fields of \textit{TransactionContext} object corresponding to the identifiers $\vec{\text{id}}$, and \( \text{exp}' = \text{exp}[\text{st}.\vec{F}(\vec{v}_{\text{s}})/ \vec{v}_{\text{s}}] \) obtained by substituting the storage variables with their corresponding object-fields in $\text{exp}$.

\begin{theorem}[Semantics Equivalence]
Let $\mathcal{S}_{\text{sol}}$ and $\mathcal{S}_{\text{java}}$ be the state transition semantic functions for Solidity and Java languages respectively. Given a Solidity state $\rho_{\text{sol}}$ and a Java state $\rho_{\text{java}}$, let $\mathbb{T}_\rho\llbracket \rho_{\text{sol}} \rrbracket = \rho_{\text{java}}$. A Solidity statement $\text{stmt}_{\text{sol}}$ is called semantically equivalent with a  Java statement $\text{stmt}_{\text{sol}}$ if $\mathcal{S}_{\text{sol}}\llbracket \text{stmt}_{\text{sol}}\rrbracket\rho_{\text{sol}} = \rho_{\text{sol}}'$, $\mathcal{S}_{\text{java}} \llbracket \text{stmt}_{\text{sol}}
\rrbracket \rho_{\text{java}} = \rho_{\text{java}}'$ and $\mathbb{T}_{\rho} \llbracket \rho_{\text{sol}}' \rrbracket = \rho_{\text{java}}'$. 

\end{theorem}
\begin{proof}
    This is proved using Floyd-Hoare logic for partial correctness. We skip the detailed proof here for brevity.
\end{proof}

\subsection{AutoJML}
This module is crucial for annotating the translated Java code with Java Modeling Language (JML) \cite{jml_language}, a foundational step for leveraging the \textit{GIN-DYN} module for loop invariant detection. Utilizing the \textit{JavaParser} library, this module systematically parses the Java source code to construct an Abstract Syntax Tree (AST) and performs a syntactic analysis of the method body, in order to extract JML preconditions and postconditions. In particular, the preconditions are determined based on parameter types, method-invoked conditions, control flow constraints, class conditions, and concurrency requirements which are present in the method-body, whereas the postconditions are framed based on return types, state changes, exception handling and external interactions. Accordingly, the methods are annotated with these pre- and postconditions, following the grammar below, where $M$ denotes Java methods. This extensive analysis allows the module to discern object state modifications and other dynamics based on the method’s contents. 

    \begin{align*}\scriptsize
        A ::= & \text{@ invariant } E; \mid \text{@ requires } E; \\
        &\mid \text{@ ensures } E; \mid \text{@ assert } E;\\
        M_a ::= & AM_a \mid M_aA \mid M\\
    \end{align*}
    
\subsection{Test Crafter}
The \textit{Test Crafter} module in our architecture is crucial for generating test cases for the code, with the objective of maximizing coverage to enrich the accuracy and comprehensiveness of resultant invariants. To this aim, we adapt the \emph{EvoSuite} \cite{evosuite} test case generator, augmenting its capabilities with custom features to focus on Solidity-specific constructs.

Monitoring gas consumption is crucial as it directly impacts the usability and cost-efficiency of smart contracts in the Ethereum network. The \textit{BlockchainSimulator} custom class within our Solidity to Java transcompiler module simulates gas consumption and estimates gas usage. We modify Evosuite to also incorporate these estimates of gas usage into fitness functions, prioritizing test scenarios which do not exceed gas limits, thereby enabling the user to ensure the economic efficiency of contract execution. Ensuring that smart contracts adhere to their declared state mutability rules is essential for maintaining the integrity and security of blockchain applications. Accordingly, we enhance EvoSuite with a new set of mutability-aware fitness functions. 

\begin{algorithm}[t]\scriptsize
\caption{Generate Solidity-Specific Test Cases}
\label{alg:testcasegeneration}
\begin{algorithmic}[1]
    \Procedure{\textit{GenTests}}{\texttt{SolCode}}
        \State $JCode \gets \text{\texttt{Translate}}(\texttt{SolCode})$
        \State $BCSim \gets \text{\texttt{InitSim}}()$
        \State \textbf{simulate} lifecycle phases: \texttt{init}, \texttt{exec}, \texttt{destruct}
        \For{\textbf{each} $annot$ \textbf{in} \texttt{JCode}}
            \State Identify and mark: \texttt{events}, \texttt{modifiers}, \texttt{address}
            \State Identify and mark \texttt{exception} handling, \texttt{transactions}
            \State \textbf{simulate} \texttt{reentrancy} for external calls
        \EndFor
        \State $EnhTests \gets \emptyset$
        \For{\textbf{each} $const$ \textbf{marked in} \texttt{JCode}}
            \State $TstSuite \gets \text{\texttt{TstGen}}(\texttt{JCode})$
            \State \textbf{integrate} \texttt{gas estimation} into fitness function
            \State \textbf{evaluate} \texttt{state mutability}, \texttt{transitions}
            \State $EnhTests \gets \texttt{EnhTests} \cup \text{\texttt{AdaptTsts}}(\texttt{TstSuite}, \texttt{const})$
            \State \textbf{check} for \texttt{state integrity} post-execution
        \EndFor
        \State \textbf{return} \texttt{EnhTests}
    \EndProcedure
\end{algorithmic}
\end{algorithm}

Different phases of smart contracts' lifecycle may significantly affect the behavior and security of blockchain applications. The transcompiler module simulates the lifecycle of a smart contract, having initialisation, execution and destruction mechanisms. We modify EvoSuite to explicitly test each phase of the smart contract using these custom mechanisms. This involves testing the initialization logic in the constructor, the execution of main functionalities, and the finality conditions under the \texttt{selfdestruct} mechanism. 

Protecting against reentrancy attacks is critical for safeguarding blockchain transactions and maintaining contract reliability. We develop a specialized component in the translator tool that simulates reentrant calls. Further, we modify Evosuite to particularly target contract functions that handle external calls or transactions. 

Proper management of exceptions is fundamental in ensuring that smart contracts react appropriately under adverse conditions. This module includes enhanced testing capabilities for Solidity’s error handling constructs like \texttt{require}, \texttt{revert}, and \texttt{assert}. Algorithm \ref{alg:testcasegeneration} outlines the algorithm of the \textit{Test Crafter}. 

\subsection{Invariant Generation Module}
\begin{figure}[t]
    \centering
    \begin{lstlisting}[language=Solidity, basicstyle=\small\ttfamily]
GasUsage <= 30000
StateAfterTxn(owner) == newOwner
AccessCheck == (msg.sender == owner)
    \end{lstlisting}
    \caption{Invariants generated from the added custom templates}
    \label{fig:invariant-examples}
\end{figure}

This module comprises of two submodules, \textit{Dynamic Invariant Explorer} and \textit{GIN-DYN}. The Dynamic Invariant Explorer module leverages Daikon \cite{daikon_tool} to generate invariants from the transcompiled Java code. Daikon is a well known dynamic invariant detection tool for Java language that analyses runtime data. Previous research, \emph{InvCon+}, has already adapted Daikon for Solidity by introducing custom invariant templates designed for financial smart contracts, integrating derivation templates like \texttt{MemberItem} and \texttt{MappingItem} for complex data structures, and the \texttt{SumMap} template to handle mappings. These enhancements help in making invariant generation process more closely aligned to Solidity-specific features. In our work we build upon these foundations to further adapt invariant detection according to the unique characteristics of Solidity and blockchain.

Estimation and management of gas consumption is highly crucial in case of Solidity smart contracts because it impacts operational costs. In this light, we introduce \texttt{GasUsage} invariant template, which ensures that the gas used by a function stays within a predefined range. This allows user to scrutinize that how operationally intensive the smart contract is. Additionally, undergoing state transitions is an inherent characteristics of smart contracts. We introduce invariant template \texttt{StateAfterTxn}, so as to capture these state changes and hence improve the overall security. Smart contracts are known to employ many access control checks, especially through modifiers, so as to constrict the execution of a function based on some validation checks. Based on this functionality, we introduce \texttt{AccessCheck} invariant template which verifies whether any access rule is there and whether only some authorized party can perform certain operation in the contract. Figure \ref{fig:invariant-examples} shows an instance of invariants generated from these added custom templates.

Additionally, in order to capture loop invariants, we adopt \textit{GIN-DYN} module, of the \textit{DynaMate} \cite{galeotti2014dynamate} which is known for its effectiveness in detecting loop invariants. In particular, \textit{GIN-DYN} module mutates post conditions and goes one step further than the pre-defined templates to generate invariants. For our tool, it proves to be exceptionally valuable for generating effective loop invariants. 

\subsection{Static Code Auditor}

The \textit{Static Code Auditor} module of the \textsf{InvSol} architecture carries out static verification of the candidate invariants produced by the \textit{Invariant Generation Module}. This is achieved through the \texttt{VerifyInvs} function depicted in Algorithm \ref{alg:intraprocess}. Figure \ref{fig:verification_procedure} shows a detailed algorithmic procedure of this function.

The primary objective of the \texttt{VerifyInvs} algorithm is to ensure that the candidate invariants hold true across all possible states of the program, thereby confirming their correctness and reliability.

The algorithm starts by initializing a set \(C_{\text{imp}}\) containing all possible logical implications between the candidate invariants in \(C_{\text{cand}}\), which are obtained in previous phase. These implications \((\alpha \Rightarrow \beta)\) suggest that if \(\alpha\) holds true, then \(\beta\) should also hold true under the same conditions. The candidate invariants themselves are referred to as \(C_{\text{cand}}\), which is the set of potential invariants identified in earlier stages of the invariant generation process.

Within the algorithm, the function \texttt{vars}(\(\alpha\)) returns the set of variables involved in an invariant \(\alpha\). This allows the algorithm to identify which program variables are being influenced or checked by the invariant. Another important function, \texttt{dep}(\(a\), \(b\)), checks for dependencies between variables \(a\) and \(b\). A dependency exists if the value of one variable affects or is influenced by the value of the other.

The core of the \texttt{VerifyInvs} algorithm is an iterative process that verifies each implication in \(C_{\text{imp}}\). For each implication \((\alpha \Rightarrow \beta)\), the algorithm examines the variables involved in both \(\alpha\) and \(\beta\) to check for any dependencies. If no such dependencies are found—meaning that the truth of \(\alpha\) does not directly influence \(\beta\)—the implication is considered valid and is removed from the set \(C_{\text{imp}}\). However, if dependencies are detected, indicating that \(\alpha\) may influence \(\beta\), the algorithm refines the candidate invariants to address these dependencies. This refinement process may involve modifying the invariants to better express the underlying relationships or to exclude invalid implications.

The algorithm continues this process iteratively, checking and refining implications until no further implications remain in \(C_{\text{imp}}\). At this point, the remaining invariants are deemed valid, as they have successfully passed through the rigorous validation process without any unresolved dependencies. These validated invariants are then finalized, providing accurate and reliable descriptions of the program’s behavior.

\begin{figure}[t]\scriptsize
\centering
\begin{tabular}{l}
\toprule
\textbf{Initialization:} \\
\quad $C_{\text{imp}} := \{ (\alpha \Rightarrow \beta) \mid \alpha, \beta \in C_{\text{cand}}, \alpha \neq \beta \}$ \\
\midrule
\textbf{Implication Verification:} \\
\textbf{While} $C_{\text{imp}} \neq \emptyset$ \textbf{do:} \\
\quad \textbf{Select} $(\alpha \Rightarrow \beta) \in C_{\text{imp}}$ \\
\quad \textbf{For all} $a \in \text{vars}(\alpha)$, $b \in \text{vars}(\beta)$, \textbf{do:} \\
\quad \quad \textbf{Check:} $\neg \text{dep}(a, b)$ \\
\quad \textbf{If} $\forall a, b$, $\neg \text{dep}(a, b)$ \textbf{holds then:} \\
\quad \quad $C_{\text{imp}} := C_{\text{imp}} \setminus \{ (\alpha \Rightarrow \beta) \}$ \\
\quad \textbf{Else:} \\
\quad \quad \textbf{Refine} $\alpha$, $\beta$ \\
\textbf{End While} \\
\bottomrule
\end{tabular}
\caption{Invariant validation by \texttt{VerifyInvs}.}
\label{fig:verification_procedure}
\end{figure}

\subsection{Specific invariant templates to address vulnerability concerns}
In \textsf{InvSol}, we introduce several advanced invariant templates to enhance the detection and prevention of vulnerabilities. These enhancements are designed to provide more adaptive and context-aware protection against the vulnerabilities. These templates are implemented by making additional enhancements to \textit{Dynamic Invariant Generator} module, ensuring automated and dynamic invariant generation without requiring user intervention.

For reentrancy, we introduce \textbf{Temporal Invariants for Read-Only Functions (TOI)}, which ensure state consistency across transactions, \textbf{Call Order Invariants (COI)} to enforce the correct sequence of function calls, and \textbf{State Transition Invariants (STI)} to verify that the contract's state transitions as expected. \textit{Dynamic Invariant Generator} implements these by capturing the initial state of relevant variables, maintaining a stack of executed functions, and modeling state transitions to ensure consistency throughout the contract's execution.

To address special storage vulnerabilities, we added \textbf{Rate of Change Invariant (RCI)}, \textbf{Contextual Limit Invariant (CLI)}, and \textbf{Dynamic Threshold Invariant (DTI)}. RCI controls the rate of change of critical state variables whereas CLI sets contextual limits based on the contract's current environment. DTI,on the other hand, applies statistical thresholds to dynamically adapt to the contract's behavior.

For dataflow vulnerabilities, we introduce \textbf{Dynamic Data Flow Threshold Invariant (DDFTI)} and \textbf{Data Flow Dependency Invariant (DFDI)}. DDFTI adjusts thresholds based on real-time data trends, while DFDI checks for consistent relationships between variables. The \textit{Invariant Generator Module} monitors the flow of data and adjusts thresholds dynamically, ensuring that relationships between data flow variables remain consistent and that more sophisticated data flow manipulations are detected.

In the context of gas control, we add \textbf{Dynamic Gas Threshold Invariant (DGTI)} and \textbf{Gas Efficiency Ratio Invariant (GERI)}. DGTI dynamically adjusts gas limits based on the function's complexity, and GERI evaluates the efficiency of gas usage, identifying abnormal patterns indicative of gas-related exploits. 

For access control, we introduce \textbf{Role-Based Access Control Invariant (RBACI)}, \textbf{Multi-Signature Access Invariant (MSAI)}, and \textbf{Ownership Transfer Validation Invariant (OTVI)}. RBACI enforces granular role-based permissions, MSAI requires multiple authorized signatures for critical actions, and OTVI ensures the legitimacy of ownership transfers. We implemente these by dynamically verifying caller roles, ensuring multiple approvals for critical actions, and validating ownership transfer processes, adding multiple layers of verification and validation. 

To address time lock vulnerabilities, we introduce three invariant templates: \textbf{Block-Time Interval Invariant (BTI)}, \textbf{Function Call Frequency Invariant (FCFI)}, and \textbf{Sequential Block Interval Invariant (SBI)}. These templates are designed to prevent rapid and unauthorized operations by enforcing temporal constraints on critical actions within the smart contract. 

To mitigate oracle slippage vulnerabilities, we introduce the \textbf{Dynamic Oracle Deviation Invariant (DODI)} and the \textbf{Price Impact Invariant (PII)}. DODI dynamically adjusts the allowed deviation in oracle-reported prices based on recent market volatility, reducing the risk of oracle manipulation during periods of high price fluctuation. PII, on the other hand, enforces a limit on the price change between consecutive oracle updates. 

For money flow vulnerabilities, we add \textbf{Cumulative Flow Tracking Invariant (CFTI)}, \textbf{Rate-Limited Flow Invariant (RLFI)}, and \textbf{Dynamic Flow Limit Invariant (DFLI)}. These templates focus on monitoring and controlling the flow of tokens within the smart contract to prevent exploitation through financial manipulations. 

Each of these invariant templates are introduced with the intention of addressing the limitations of static and rigid checks, providing \textsf{InvSol} with the ability to dynamically adapt to changing conditions and detect more complex attack vectors. Table \ref{tab:invsol_invariants} gives an overveiw of these templates.

\begin{table*}[t]
\centering
\caption{Specific Invariant Templates Added in \textsf{InvSol}}
\scriptsize
\begin{tabular}{|p{2.0cm}|p{3.0cm}|p{1.0cm}|p{8.0cm}|}
\hline
\textbf{Vulnerability} & \textbf{Invariant Templates} & \textbf{Short Form} & \textbf{Purpose} \\ \hline

\multirow{3}{2.0cm}{\textbf{Reentrancy}} 
& Temporal Invariants for Read-Only Functions & \textbf{TOI} & Ensures state consistency in read-only functions across transactions. \\ \cline{2-4}
& Call Order Invariants & \textbf{COI} & Verifies correct sequence of function calls to prevent reentrancy exploits. \\ \cline{2-4}
& State Transition Invariants & \textbf{STI} & Confirms proper state transitions during function execution. \\ \hline

\multirow{3}{2.0cm}{\textbf{Special Storage}} 
& Rate of Change Invariant & \textbf{RCI} & Controls the rate at which critical storage variables can change. \\ \cline{2-4}
& Contextual Limit Invariant & \textbf{CLI} & Sets limits based on the contract's current state or external factors. \\ \cline{2-4}
& Dynamic Threshold Invariant & \textbf{DTI} & Applies statistical thresholds to critical variables based on historical data. \\ \hline

\multirow{2}{2.0cm}{\textbf{Dataflow}} 
& Dynamic Data Flow Threshold Invariant & \textbf{DDFTI} & Adapts thresholds for data flows to catch anomalies during execution. \\ \cline{2-4}
& Data Flow Dependency Invariant & \textbf{DFDI} & Ensures consistent relationships between related data flow variables. \\ \hline

\multirow{2}{2.0cm}{\textbf{Gas Control}} 
& Dynamic Gas Threshold Invariant & \textbf{DGTI} & Sets adaptive gas limits based on function complexity and historical usage. \\ \cline{2-4}
& Gas Efficiency Ratio Invariant & \textbf{GERI} & Monitors gas usage efficiency to detect abnormal consumption patterns. \\ \hline

\multirow{3}{2.0cm}{\textbf{Access Control}} 
& Role-Based Access Control Invariant & \textbf{RBACI} & Ensures granular access control by defining roles and associated permissions. \\ \cline{2-4}
& Multi-Signature Access Invariant & \textbf{MSAI} & Requires multiple signatures for critical function execution. \\ \cline{2-4}
& Ownership Transfer Validation Invariant & \textbf{OTVI} & Validates ownership transfers to prevent unauthorized changes. \\ \hline

\multirow{3}{2.0cm}{\textbf{Time Lock}} 
& Block-Time Interval Invariant & \textbf{BTI} & Ensures a minimum time interval between critical operations to prevent rapid exploits. \\ \cline{2-4}
& Function Call Frequency Invariant & \textbf{FCFI} & Limits the number of times a function can be called within a specific time unit. \\ \cline{2-4}
& Sequential Block Interval Invariant & \textbf{SBI} & Ensures a minimum block interval between critical operations to prevent rapid consecutive actions. \\ \hline

\multirow{2}{2.0cm}{\textbf{Oracle Slippage}} 
& Dynamic Oracle Deviation Invariant & \textbf{DODI} & Adjusts the allowed price deviation dynamically based on recent volatility to prevent price manipulation. \\ \cline{2-4}
& Price Impact Invariant & \textbf{PII} & Enforces a limit on the price change that can occur between consecutive oracle updates to prevent large, sudden shifts. \\ \hline

\multirow{3}{2.0cm}{\textbf{Money Flow}} 
& Cumulative Flow Tracking Invariant & \textbf{CFTI} & Tracks cumulative token flow over a specified period to prevent gradual exploitation. \\ \cline{2-4}
& Rate-Limited Flow Invariant & \textbf{RLFI} & Limits the rate of token flow within a certain timeframe to prevent rapid large transactions. \\ \cline{2-4}
& Dynamic Flow Limit Invariant & \textbf{DFLI} & Sets dynamic flow limits based on contract state. \\ \hline

\end{tabular}
\label{tab:invsol_invariants}
\end{table*}

\section{Experimental Evaluation}
\label{label:experimental-evaluation}

\begin{table*}[ht]
\centering
\caption{Overview of Ethereum ETL dataset}
\label{tab:erc_overview}
\renewcommand{\arraystretch}{1.2}
\scriptsize
\begin{tabularx}{\textwidth}{|p{1.2cm}|X|X|p{1.2cm}|p{0.8cm}|}
\hline
\textbf{Contract Types} & \textbf{Characteristics} & \textbf{Common Use Cases} & \textbf{Solidity Versions} & \textbf{\#SC} \\ \hline
\textbf{ERC20} & Fungible tokens. Standard functions include \texttt{totalSupply}, \texttt{balanceOf}, \texttt{transfer}, \texttt{approve}, \texttt{transferFrom}. Widely supported. & Cryptocurrencies, utility tokens, governance tokens, stablecoins, DeFi protocols. & 0.4.17+ & 211 \\ \hline
\textbf{ERC721} & Non-fungible tokens. Unique assets. Standard functions include \texttt{ownerOf}, \texttt{transferFrom}, \texttt{approve}, \texttt{getApproved}. Metadata extensions. & Digital collectibles, gaming assets, digital art, virtual real estate, certificates. & 0.4.24+ & 103 \\ \hline
\textbf{ERC1155} & Multi-token standard. Supports both fungible and non-fungible tokens. Batch operations include \texttt{safeTransferFrom}, \texttt{safeBatchTransferFrom}. & Gaming assets, digital art, tokenized real-world assets. & 0.5.0+ & 52 \\ \hline
\textbf{ERC777} & Fungible tokens. Advanced features. Backward compatible with ERC20. Hooks for send and receive functions. & Cryptocurrencies, DeFi protocols, reward tokens. & 0.5.0+ & 33 \\ \hline
\textbf{Others} & Various specialized functionalities such as Crowdsale, DAO, Staking, Voting, Escrow, Insurance, Oracle, Payment Channel, DeFi, Identity, Supply Chain, Lottery, Registry, Time-Locked operations. & Fundraising, governance, rewards, secure transactions, insurance, data fetching, off-chain transactions, financial applications, digital identities, supply chain tracking, gaming, asset registries, time-locked operations. & Varies & 94 \\ \hline
\end{tabularx}
\end{table*}

\begin{table}[ht]
\centering\scriptsize
\caption{Ethereum ETL dataset statistics according to LoC and Category}
\label{tab:loc_ranges}
\resizebox{\columnwidth}{!}{\begin{tabular}{|c|c|c|c|c|c|c|}
\hline
\textbf{LOC Range} & \textbf{ERC20} & \textbf{ERC721} & \textbf{ERC1155} & \textbf{ERC777} & \textbf{Others} & \textbf{Total} \\ \hline
0-100   & 22  & 12  & 6   & 3   & 7   & 50  \\ \hline
101-200 & 90  & 45  & 20  & 12  & 33  & 200 \\ \hline
201-300 & 9   & 4   & 2   & 1   & 4   & 20  \\ \hline
301+    & 90  & 42  & 24  & 17  & 50  & 223 \\ \hline
\textbf{Total} & \textbf{211} & \textbf{103} & \textbf{52} & \textbf{33} & \textbf{94} & \textbf{493} \\ \hline
\end{tabular}}
\end{table}

% Add this line inside the document environment where you want the table to appear
\begin{table}[t]
\centering
\scriptsize
\caption{Overview of SmartBugs Dataset}
\resizebox{\columnwidth}{!}{\begin{tabular}{|l|c|c|c|c|}
\hline
\multirow{2}{*}{\textbf{Vulnerability}} & \multirow{2}{*}{\textbf{\#SC}} & \multicolumn{3}{c|}{\textbf{LoC}}\\ \cline{3-5}
& & \textbf{$\textless$ 200} & \textbf{200-1000} & \textbf{$\textgreater$ 1000} \\
\hline
Access Control & 18 & 16 & 2 & 0 \\
Arithmetic & 15 & 14 & 1 & 0 \\
Bad Randomness & 8 & 6 & 2 & 0 \\
Denial of Service & 6 & 6 & 0 & 0 \\
Front Running & 4 & 4 & 0 & 0 \\
Other & 3 & 3 & 0 & 0 \\
Reentrancy & 31 & 30 & 1 & 0 \\
Short Addresses & 1 & 1 & 0 & 0 \\
Time Manipulation & 5 & 5 & 0 & 0 \\
Unchecked Low Level Calls & 52 & 46 & 5 & 1 \\

\hline
\end{tabular}}
\label{table:smartbugs}
\end{table}

%\subsection{Research Questions}
In this section, we perform a comprehensive evaluation of \textsf{InvSol} on two well-established benchmark datasets and we assess its efficacy by addressing the following four key research questions:
\begin{itemize}
    \item RQ1: \textit{Invariant detection prior to smart contract deployment.} How does \textsf{InvSol} perform with respect to current state-of-the-art approaches?
    \item RQ2: \textit{Loop Invariants.} How does the proficiency of state-of-the-art tools compare with \textsf{InvSol} in identifying invariants in presence of diverse Solidity constructs and loop invariants?
    \item RQ3: \textit{Impact of scarcity of transaction history on quality of invariants.} Can \textsf{InvSol} offer a superior performance than state-of-the-art tools for young smart contracts?
    \item RQ4: \textit{Enhancing the Security of Smart Contracts.} To what extent does \textsf{InvSol} help in securing the real world smart contracts against the well known vulnerabilities?
    \item RQ5: \textit{Comparison of invariant effectiveness across key vulnerabilities.} How does \textsf{InvSol} compare with \emph{Trace2Inv} \cite{chen2024demystifying} in addressing critical vulnerabilities.

\end{itemize}

%\cite{fraser2012sound}

\subsection{Benchmark Datasets}\label{subsec:dataset}
We conduct experiments on two benchmark datasets, sourced from \emph{Google BigQuery} platform and \emph{SmartBugs}, covering a total of 678 real world solidity contracts. Let us briefly describe them below:

\begin{itemize}
    \item \emph{Google BigQuery} \cite{google_bigquery} provides a user friendly platform to query large datasets. Aiming to demonstrate the ability of \textsf{InvSol} compared to the state-of-the-art approaches on a diverse set of Solidity constructs including loops, we used the public Ethereum ETL dataset available on the \emph{Google BigQuery} platform. We query to collect Solidity smart contracts deployed between 01-01-2022 to 01-01-2023. We choose this time frame to capture all the trends, features and programming advancements that have been prevalent in Solidity language lately. From the collected smart contracts, we select a set of 456 smart contracts covering different varieties and applications such as cryptocurrencies, DeFi protocols, governance tokens, digital collectibles, gaming assets, and real estate tokens, in order to ensure a rigorous evaluation of our tool. Additionally, in order to address the RQ3, we selected 37 newly deployed smart contracts from \emph{BigQuery} which have less than 50 historical transactions. The detailed statistics of these datasets are depicted in Tables \ref{tab:erc_overview} and \ref{tab:loc_ranges} respectively. 

    \item The \emph{SmartBugs Curated} \cite{smartbugs_dataset} datatset is a well-known benchmark collection of vulnerable smart contracts, widely recognized for its comprehensive coverage of real world vulnerabilities. We test \textsf{InvSol} on all 143 smart contracts present in this dataset in order to find out how our tool performs in detecting or helping in mitigating these known vulnerabilities. Table \ref{table:smartbugs} provides a comprehensive overview of this dataset.

    \item The \emph{Trace2Inv} \cite{chen2024demystifying} benchmark dataset comprises 42 real-world smart contracts that were exploited in security attacks, leading to significant financial losses on the Ethereum blockchain. Collected between February 2020 and August 2022, this dataset serves as a critical benchmark for evaluating the effectiveness of invariant detection tool \emph{Trace2Inv}.

\end{itemize}
All experiments are conducted on a dedicated Ubuntu platform, which is configured with a 1.6 GHz Intel Core i5 processor and equipped with 8 GB of RAM. We focus on four specific classes of invariants, namely state invariants, arithematic invariants, boundary invariants and security invariants, detailed in Table \ref{tab:invariants_types}. We manually verify the generated invariants for correctness and relevance. We take help from well known external resources such as official Solidity Documentation for syntax and best practices, OpenZeppelin Contracts \cite{openzeppelin_contracts} for standardized and audited smart contract implementations, Ethereum Improvement Proposals (EIPs) \cite{ethereum_eip} for the latest standards and enhancements, and the SWC Registry (Smart Contract Weakness Classification) \cite{swc_registry} for addressing known vulnerabilities and security best practices. 

\begin{table*}[t]
\centering
\caption{Description of Invariant Classes of Solidity Smart Contracts}
\label{tab:invariants_types}
\begin{tabularx}{\textwidth}{@{}l>{\hsize=.7\hsize}YY@{}}
\toprule
Invariant Classes & Description & Example Templates \\ \midrule
State Invariants & Check for unchanging state properties between function calls. & \begin{itemize}[leftmargin=*, nosep, before=\vspace{-\baselineskip}, after=\vspace{-\baselineskip}]
    \item \texttt{this.balance == orig(this.balance)}
    \item \texttt{owner == orig(owner)}
\end{itemize} \\
Arithmetic Invariants & Concerned with mathematical relationships, ensuring computations do not cause errors like overflow. & \begin{itemize}[leftmargin=*, nosep, before=\vspace{-\baselineskip}, after=\vspace{-\baselineskip}]
    \item \texttt{x + y >= x}
    \item \texttt{totalSupply == sum(balances[])}
\end{itemize} \\
Boundary Invariants & Ensure variables stay within predefined limits, crucial for avoiding logical errors. & \begin{itemize}[leftmargin=*, nosep, before=\vspace{-\baselineskip}, after=\vspace{-\baselineskip}]
    \item \texttt{msg.value > 0}
    \item \texttt{tokensIssued <= maxTokens}
\end{itemize} \\
Security Invariants & Focus on conditions that safeguard against vulnerabilities specific to smart contracts. & \begin{itemize}[leftmargin=*, nosep, before=\vspace{-\baselineskip}, after=\vspace{-\baselineskip}]
    \item \texttt{msg.sender == owner}
    \item \texttt{state == State.CREATED}
\end{itemize} \\
\bottomrule
\end{tabularx}
\end{table*}

\begin{table}[t]
\centering
\large % You can also try \normalsize if \large is too big
\caption{Performance Comparison}
\label{tab:invariants_comparison}
\resizebox{\columnwidth}{!}{ % Adjusted to fit one full column width
\begin{tabular}{|l|c|c|c|c|c|c|}
\hline
\textbf{Type} & \multicolumn{2}{c|}{\textbf{InvSol}} & \multicolumn{2}{c|}{\textbf{InvCon+}} & \multicolumn{2}{c|}{\textbf{InvCon}} \\ \hline
 & \textbf{True Pos.} & \textbf{Precision} & \textbf{True Pos.} & \textbf{Precision} & \textbf{True Pos.} & \textbf{Precision} \\ \hline
\textbf{State Invariants}      & 6340  & 100\%  & 6023  & 100\%  & 5031  & 9.3\% \\ \hline
\textbf{Arithmetic Invariants} & 5706  & 100\%  & 5568  & 97.6\% & 5193  & 9.7\% \\ \hline
\textbf{Boundary Invariants}   & 3804  & 100\%  & 3977  & 100\%  & 3361  & 8.7\% \\ \hline
\textbf{Security Invariants}   & 2510  & 98.5\% & 2421  & 100\%  & 3676  & 9.3\% \\ \hline
\end{tabular}}
\end{table}

\begin{table}[t]
\centering
\large % You can also try \normalsize if \large is too big
\caption{Performance of \textsf{InvSol} for loop invariant}
\label{tab:loop_invariants}
%\small
\resizebox{\columnwidth}{!}{\begin{tabular}{|l|c|c|c|c|c|c|}
\hline
Category       & \# Contracts & State Inv. & Arith. Inv. & Bound. Inv. & Sec. Inv. & Precision \\ \hline
Simple Loops   & 6            & 15         & 14          & 9           & 6         & 100\%     \\ \hline
Nested Loops   & 13           & 36         & 32          & 22          & 14        & 98\%      \\ \hline
Dynamic Loops  & 16           & 41         & 38          & 25          & 17        & 93\%      \\ \hline
\end{tabular}}
\end{table}

\subsubsection{Experiment for RQ1} 

To address Research Question 1 (RQ1), we evaluate \textsf{InvSol} alongside the existing tools \emph{InvCon+} and \emph{InvCon} on 634 smart contracts belonging to the first two datasets. Table \ref{tab:invariants_comparison} illustrates \textsf{InvSol}'s performance compared to these state-of-the-art tools in generating various classes of invariants. Notably, although \textsf{InvSol} does not utilize real-world transaction history, its performance is comparable to that of \emph{InvCon+}. While \emph{InvCon} generates a large number of invariants, it falls short in precision. 

\begin{center}
\noindent\colorbox{lightgray}{%
    \parbox{0.95\columnwidth}{%
        \textit{Finding 1:} \textsf{InvSol} outperforms state-of-the-art tools in precision, demonstrating its robustness in invariant generation without relying on historical transaction data.
    }%
}
\end{center}

\subsubsection{Experiment for RQ2} 
%As discussed earlier, currently no research work in this domain addresses loop invariants in Solidity. 
To address RQ2, we select from Ethereum ETL dataset a set of 35 smart contracts which contain various loops structures. For critical analysis of the results, we classify these loop structures into following three categories: simple, nested and dynamic. Simple loops in Solidity are straightforward iterations over fixed ranges, often used to aggregate or compute sums of blockchain-based tokens or data. Nested loops are loops which involve multiple looping mechanisms one within the other, and are crucial for handling more complex data structures or transactional processes. Dynamic loops, on the other hand, are characterized by variable termination conditions which may depend on runtime data or external blockchain events. The evaluation results are depicted in Table \ref{tab:loop_invariants}, which show that \textsf{InvSol} is able to generate precise loop invariants, for the loops of all the categories. For simple loops, \textsf{InvSol} achieves a perfect precision of 100\%, detecting 44 invariants with a balanced distribution of state, arithmetic, boundary, and security invariants. In nested loops, it maintains a high precision of 98\%, identifying 104 invariants. In case of Dynamic loops, which are inherently more complex, \textsf{InvSol} achieves a precision of 93\% with 121 detected invariants. 

\begin{center}
\noindent\colorbox{lightgray}{%
    \parbox{0.95\columnwidth}{%
        \textit{Finding 2:} The ability of \textsf{InvSol} to generate precise loop invariants across various Solidity constructs, including dynamic loops, reveals a significant advancement in handling the intrinsic complexity of smart contracts. This proficiency not only underscores \textsf{InvSol}'s robustness but also suggests its potential to redefine the standard for loop invariant detection, ensuring higher reliability and security in contract verification processes.
    }%
}
\end{center}

\subsubsection{Experiment for RQ3} 
For Research Question 3 (RQ3), we perform experiment on 37 newly deployed smart contracts present in Ethereum ETL dataset. We do this so as to test \textsf{InvSol} in realistic scenarios where smart contracts are newly deployed or used less frequently, with limited transaction data. As shown in Table \ref{tab:performance_summary}, in case of scarce transaction history, \textsf{InvSol} clearly outperforms \emph{InvCon+} and \emph{InvCon} across all invariant types, achieving high precision and recall rates. These results clearly indicate that while \emph{InvCon+} and \emph{InvCon} struggle with limited transaction history, \textsf{InvSol}, which does not depend on transaction history for invariant detection, performs effectively in such scenarios.

\begin{table}[t]
\centering
\scriptsize
\caption{Performance Summary with Scarce Transaction History}
\label{tab:performance_summary}
\resizebox{\columnwidth}{!}{%
\begin{tabular}{|l|c|c|c|c|c|c|c|c|}
\hline
\textbf{Tool} & \multicolumn{2}{c|}{\textbf{State Inv.}} & \multicolumn{2}{c|}{\textbf{Arith. Inv.}} & \multicolumn{2}{c|}{\textbf{Boundary Inv.}} & \multicolumn{2}{c|}{\textbf{Security Inv.}} \\
\hline
& \textbf{Prec.} & \textbf{Rec.} & \textbf{Prec.} & \textbf{Rec.} & \textbf{Prec.} & \textbf{Rec.} & \textbf{Prec.} & \textbf{Rec.} \\
\hline
\textbf{InvSol}  & 98\%  & 80\% & 97\%  & 78\% & 99\%  & 79\% & 98\%  & 79\% \\ \hline
\textbf{InvCon+} & 24\%  & 12\% & 26\%  & 10\% & 25\%  & 11\% & 25\%  & 11\% \\ \hline
\textbf{InvCon}  & 5\%  & 4\%  & 8\%   & 4\%  & 7\%   & 0\%  & 5\%   & 1\%  \\ \hline
\end{tabular}
}
\end{table}

\begin{figure}[t]
    \centering
    \begin{tikzpicture}
        \begin{axis}[
            width=0.9\columnwidth,
            height=0.2\textheight,
            xlabel={\small\text{\#Transactions}},
            ylabel={\small\text{Recall Score (\%)}},
            xmin=0, xmax=500,
            ymin=0, ymax=100,
            xtick={0, 50, 100, 150, 200, 250, 300, 350, 400, 450, 500},
            ytick={0, 25, 50, 75, 100},
            tick label style={font=\scriptsize}, % Reduce the size of axis labels
            legend style={at={(0.98,0.475)}, anchor=east, font=\scriptsize}, % Move legend to the extreme right
            grid=major
        ]
            % Data for InvSol
            \addplot[color=blue, mark=*] coordinates {
                (0, 78) (50, 79) (100, 81) (150, 80) (200, 82) (250, 78) (300, 80) (350, 81) (400, 79) (450, 80) (500, 82)
            };
            \addlegendentry{InvSol}

            % Data for InvCon+
            \addplot[color=orange, mark=square*] coordinates {
                (0, 0) (50, 15) (100, 30) (150, 45) (200, 50) (250, 60) (300, 65) (350, 70) (400, 75) (450, 78) (500, 80)
            };
            \addlegendentry{InvCon+}
            
            % Data for InvCon
            \addplot[color={rgb,255:red,255;green,204;blue,0}, mark=triangle*] coordinates {
                (0, 0) (50, 2) (100, 5) (150, 8) (200, 10) (250, 12) (300, 14) (350, 16) (400, 18) (450, 19) (500, 20)
            };
            \addlegendentry{InvCon}
        \end{axis}
    \end{tikzpicture}
    \caption{Average recall score vs. increasing number of transactions}
    \label{fig:recall_over_transactions}
\end{figure}

This fact becomes more evident in Figure \ref{fig:recall_over_transactions}, which compares the average recall scores achieved by \textsf{InvSol}, \emph{InvCon+}, and \emph{InvCon} as the number of transactions increases. As observed, the recall score for \textsf{InvSol} remains consistently high, fluctuating slightly between 77\% and 83\%, regardless of the number of transactions. In contrast, \emph{InvCon+} starts with a recall score of 0\% with no transactions and gradually increases to approximately 80\% as the transaction count reaches 500. \emph{InvCon}, on the other hand, shows a much slower increase in recall score, only reaching around 20\% after 500 transactions. This highlights the robustness of \textsf{InvSol} in maintaining high recall scores even with limited transaction history, whereas \emph{InvCon+} and \emph{InvCon} require a significant number of transactions to achieve similar levels of performance.

\begin{center}
\noindent\colorbox{lightgray}{%
    \parbox{0.95\columnwidth}{%
        \textit{Finding 3:} \textsf{InvSol} demonstrates superior performance in invariant detection with minimal transaction history, proving its robustness for early-stage smart contract verification.
    }%
}
\end{center}

\subsubsection{Experiment for RQ4.}

\begin{table*}[t]
\centering\scriptsize
\caption{Effectiveness of Invariant Classes Against Specific Vulnerabilities}
\label{tab:invariant_effectiveness}
\resizebox{\textwidth}{!}{%
\begin{tabular}{|l|c|c|c|c|c|}
\hline
\textbf{Vulnerability} & \textbf{State Invariants} & \textbf{Arithmetic Invariants} & \textbf{Boundary Invariants} & \textbf{Security Invariants} & \textbf{Loop Invariants} \\ \hline
Reentrancy & \checkmark & \texttimes & \texttimes & \checkmark & \texttimes \\ \hline
Access Control & \texttimes & \texttimes & \texttimes & \checkmark & \texttimes \\ \hline
Arithmetic & \texttimes & \checkmark & \checkmark & \texttimes & \texttimes \\ \hline
Unchecked Low Level Calls & \checkmark & \texttimes & \texttimes & \texttimes & \texttimes \\ \hline
Denial of Service & \checkmark & \texttimes & \checkmark & \texttimes & \texttimes \\ \hline
Bad Randomness & \texttimes & \texttimes & \texttimes & \texttimes & \texttimes \\ \hline
Front Running & \texttimes & \texttimes & \texttimes & \texttimes & \texttimes \\ \hline
Time Manipulation & \checkmark & \texttimes & \texttimes & \texttimes & \texttimes \\ \hline
Short Addresses & \texttimes & \texttimes & \checkmark & \texttimes & \texttimes \\ \hline
\end{tabular}}
\end{table*}

\begin{table}[ht]
\centering
\scriptsize
\caption{Performance comparison for vulnerability detection}
\label{tab:vulnerability_evaluation}
\resizebox{\columnwidth}{!}{\begin{tabular}{|l|c|c|c|}
\hline
\textbf{Vulnerability} & \textbf{InvCon (\%)} & \textbf{InvCon+ (\%)} & \textbf{InvSol (\%)} \\
\hline
Access Control & 55.81 & 62.04 & 69.63 \\
Arithmetic & 51.81 & 59.51 & 76.73 \\
Bad Randomness & 60.37 & 67.30 & 87.66 \\
Denial of Service & 55.00 & 63.35 & 72.34 \\
Front Running & 45.15 & 52.87 & 61.12 \\
Reentrancy & 0 & 0 & 0 \\
Short Addresses & 95.87 & 100 & 100 \\
Time Manipulation & 0 & 0 & 0 \\
Unchecked Calls & 60.27 & 67.10 & 88.34 \\
\hline
\end{tabular}}
\end{table}

As discussed earlier, \textsf{InvSol} generates invariants before a smart contract is deployed, while existing tools depend on the transaction history of contracts that are already deployed. Because blockchain is immutable, relying on post-deployment data to generate invariants can make smart contracts vulnerable. If these important checks are only made after the contract is live, especially in financial contexts, it might be too late to prevent any potential financial losses.

Therefore, the aim of this experiment is to evaluate the effectiveness of \textsf{InvSol} in detecting the presence of vulnerabilities in smart contracts prior to  their deployment. In order to provide a fair ground of comparison, the evaluation is performed based on the test cases generated by the Evosuite test case generator for all the tools. Table \ref{tab:invariant_effectiveness} offers a detailed overview of which types of invariants are most effective for addressing each specific vulnerability. The evaluation results, based on both \emph{SmartBugs Curated} dataset with nine vulnerabilites are presented in Table \ref{tab:vulnerability_evaluation}. This results reveal significant performance differences between \textsf{InvSol}, \emph{InvCon} and \emph{InvCon+} across various types of vulnerabilities. \textsf{InvSol} consistently outperforms \emph{InvCon} and \emph{InvCon+} in detecting vulnerabilities such as Arithmetic, Bad Randomness, Unchecked Calls, Access Control, and Denial of Service. However, both tools show limitations in addressing Time Manipulation and Reentrancy vulnerabilities. We further noticed that in case of Off by One errors also, \textsf{InvSol} showed effectiveness while the other tools failed to give any indication of this vulnerability. 

\textsf{InvSol's} success in addressing Arithematic Vulnerabilities is attributed to 
integration of custom JML annotations that impose strict preconditions, specifying exact variable ranges required before arithmetic operations. Postconditions are also added to check that results remain within clearly defined safe limits. Furthermore, the \textit{Test Crafter} module is enhanced to systematically generate and explore edge cases, targeting scenarios that might trigger overflow or underflow. This combined approach ensures that any potential arithmetic overflow or underflow is precisely flagged, thus increasing the tool’s accuracy.

\textsf{InvSol} identifies potential risk of Denial of Service by implementing static gas cost estimation, during test execution. Furthermore, loop bound analysis incorporated using JML annotations and the \textit{Test Crafter Module} helps in Denial of Service risk detection. \textsf{InvSol's} tries to simulate multiple sequential transactions with varying gas prices and submission timings, in order to address Front Running vulnerabilities, but this functionality needs further refinement for optimal performance.

\begin{center}
\noindent\colorbox{lightgray}{%
    \parbox{0.95\columnwidth}{%
        \textit{Finding 4:} \textsf{InvSol} shifts the paradigm in smart contract security by effectively preempting vulnerabilities before deployment, particularly excelling in detecting critical issues like arithmetic errors and unchecked calls. This proactive approach not only reduces the risk of financial losses but also challenges the reliance on reactive measures, setting a new standard for contract robustness in dynamic blockchain environments.
    }%
}
\end{center}

\subsubsection{Experiment for RQ5}
The authors in \cite{chen2024demystifying} presents a comprehensive tool, namely \textsf{Trace2Inv}, which dynamically generates and evaluates the effectiveness of invariants based on transaction history to safeguard the smart contracts. In this section, we evaluate \textsf{InvSol} w.r.t \textsf{Trace2Inv} on the same benchmark dataset as used in \cite{chen2024demystifying}, considering the following two data settings: (1) dynamic transaction data generated from the Ethereum main net and (2) static data generated by our proposed \textit{TestCrafter}. The corresponding results are depicted in Tables \ref{tab:invsol_vs_trace2inv_1} and \ref{tab:invsol_vs_trace2inv_2} respectively. Interestingly, we observe that, in both the cases, \textsf{InvSol} is performing better than \textsf{Trace2Inv}. 

The results from the two tables show that \textsf{InvSol} outperforms \textsf{Trace2Inv} in protecting smart contracts and blocking vulnerabilities across various categories. For reentrancy vulnerabilities, \textsf{InvSol}'s templates like Temporal Order Invariants (TOI) and Call Order Invariants (COI) offer comparable or slightly better protection, especially in blocking attacks. In the case of special storage vulnerabilities, \textsf{InvSol}'s Rate of Change Invariant (RCI) protects more contracts (7 out of 42) and blocks more exploits (5 out of 27) compared to \textsf{Trace2Inv}'s TotalSupplyUpperBound (TSU) and TotalBorrowUpperBound (TBU). For gas control, \textsf{InvSol}'s Dynamic Gas Threshold Invariant (DGTI) protects more contracts and blocks more exploits than \textsf{Trace2Inv}'s GS and GC invariants. Finally, in access control, \textsf{InvSol}'s Role-Based Access Control Invariant (RBACI) and Multi-Signature Access Invariant (MSAI) show superior performance by protecting more contracts (9 out of 42) and blocking more exploits compared to \textsf{Trace2Inv}'s access control invariants. 

Even though our templates fall short in case of Time Lock, Oracle Slippage and Money Flow, yet the are not much far behind and are able to detect crucial invariants related to these vulnerabilities in some smart contracts.

\begin{table}[h]
\centering\scriptsize
\begin{tabular}{|p{2.0cm}|p{2.5cm}|p{1.0cm}|p{1.0cm}|}
\hline
\textbf{Vulnerability} & \textbf{Tool} & \textbf{\rotatebox{90}{\# Protected (42)}} & \textbf{\rotatebox{90}{\# Blocked (27)}} \\ \hline

\multirow{4}{2.0cm}{\textbf{Reentrancy}} 
& \textbf{Trace2Inv (RE)} & 2 & 2 \\ \cline{2-4}
& \textbf{InvSol (TOI)} & 2 & 3 \\ \cline{2-4} 
& \textbf{InvSol (COI)} & 2 & 2 \\ \cline{2-4}
& \textbf{InvSol (STI)} & 2 & 1 \\ \hline

\multirow{5}{2.0cm}{\textbf{Special Storage}} 
& \textbf{Trace2Inv (TSU)} & 5 & 4 \\ \cline{2-4}
& \textbf{Trace2Inv (TBU)} & 3 & 3 \\ \cline{2-4}
& \textbf{InvSol (RCI)} & 7 & 5 \\ \cline{2-4} 
& \textbf{InvSol (CLI)} & 4 & 3 \\ \cline{2-4}
& \textbf{InvSol (DTI)} & 3 & 2 \\ \hline

\multirow{6}{2.0cm}{\textbf{Dataflow}} 
& \textbf{Trace2Inv (MU)} & 3 & 2 \\ \cline{2-4}
& \textbf{Trace2Inv (CVU)} & 2 & 1 \\ \cline{2-4}
& \textbf{Trace2Inv (DFU)} & 1 & 1 \\ \cline{2-4}
& \textbf{Trace2Inv (DFL)} & 1 & 1 \\ \cline{2-4}
& \textbf{InvSol (DDFTI)} & 4 & 4 \\ \cline{2-4} 
& \textbf{InvSol (DFDI)} & 4 & 2 \\ \hline 

\multirow{4}{2.0cm}{\textbf{Gas Control}} 
& \textbf{Trace2Inv (GS)} & 8 & 7 \\ \cline{2-4}
& \textbf{Trace2Inv (GC)} & 7 & 6 \\ \cline{2-4}
& \textbf{InvSol (DGTI)} & 9 & 8 \\ \cline{2-4}
& \textbf{InvSol (GERI)} & 8 & 7 \\ \hline

\multirow{7}{2.0cm}{\textbf{Access Control}} 
& \textbf{Trace2Inv (EOA)} & 7 & 4 \\ \cline{2-4}
& \textbf{Trace2Inv (SO)} & 6 & 4 \\ \cline{2-4}
& \textbf{Trace2Inv (OO)} & 4 & 3 \\ \cline{2-4}
& \textbf{Trace2Inv (SM)} & 4 & 2 \\ \cline{2-4}
& \textbf{Trace2Inv (OM)} & 2 & 2 \\ \cline{2-4}
& \textbf{InvSol (RBACI)} & 9 & 6 \\ \cline{2-4}
& \textbf{InvSol (MSAI)} & 8 & 6 \\ \cline{2-4}
& \textbf{InvSol (OTVI)} & 7 & 5 \\ \hline

\multirow{3}{2.0cm}{\textbf{Time Lock}} 
& \textbf{Trace2Inv (SB)} & 5 & 3 \\ \cline{2-4}
& \textbf{Trace2Inv (OB)} & 4 & 2 \\ \cline{2-4}
& \textbf{Trace2Inv (LU)} & 3 & 2 \\ \cline{2-4}
& \textbf{InvSol (BTI)} & 2 & 2 \\ \cline{2-4} 
& \textbf{InvSol (FCFI)} & 2 & 1 \\ \cline{2-4}
& \textbf{InvSol (SBI)} & 2 & 1 \\ \hline

\multirow{3}{2.0cm}{\textbf{Oracle Slippage}} 
& \textbf{Trace2Inv (OR)} & 6 & 4 \\ \cline{2-4}
& \textbf{Trace2Inv (OD)} & 5 & 3 \\ \cline{2-4}
& \textbf{InvSol (DODI)} & 4 & 2 \\ \cline{2-4}
& \textbf{InvSol (PII)} & 4 & 2 \\ \hline

\multirow{4}{2.0cm}{\textbf{Money Flow}} 
& \textbf{Trace2Inv (TIU)} & 7 & 5 \\ \cline{2-4}
& \textbf{Trace2Inv (TOU)} & 6 & 4 \\ \cline{2-4} 
& \textbf{Trace2Inv (TIRU)} & 5 & 3 \\ \cline{2-4}
& \textbf{Trace2Inv (TORU)} & 4 & 2 \\ \cline{2-4}
& \textbf{InvSol (CFTI)} & 3 & 3 \\ \cline{2-4} 
& \textbf{InvSol (RLFI)} & 3 & 3 \\ \cline{2-4}
& \textbf{InvSol (DFLI)} & 3 & 2 \\ \hline

\end{tabular}
\caption{InvSol Vs. Trace2Inv on historical transaction data}
\label{tab:invsol_vs_trace2inv_1}
\end{table}

\begin{table}[h]
\centering\scriptsize
\begin{tabular}{|p{2.0cm}|p{2.5cm}|p{1.0cm}|p{1.0cm}|}
\hline
\textbf{Vulnerability} & \textbf{Tool} & \textbf{\rotatebox{90}{\# Protected (42)}} & \textbf{\rotatebox{90}{\# Blocked (27)}} \\ \hline

\multirow{4}{2.0cm}{\textbf{Reentrancy}} 
& \textbf{Trace2Inv (RE)} & 2 & 2 \\ \cline{2-4}
& \textbf{InvSol (TOI)} & 2 & 2 \\ \cline{2-4} 
& \textbf{InvSol (COI)} & 2 & 2 \\ \cline{2-4}
& \textbf{InvSol (STI)} & 2 & 1 \\ \hline

\multirow{5}{2.0cm}{\textbf{Special Storage}} 
& \textbf{Trace2Inv (TSU)} & 5 & 4 \\ \cline{2-4}
& \textbf{Trace2Inv (TBU)} & 3 & 4 \\ \cline{2-4} 
& \textbf{InvSol (RCI)} & 7 & 5 \\ \cline{2-4} 
& \textbf{InvSol (CLI)} & 4 & 3 \\ \cline{2-4}
& \textbf{InvSol (DTI)} & 3 & 2 \\ \hline

\multirow{6}{2.0cm}{\textbf{Dataflow}} 
& \textbf{Trace2Inv (MU)} & 3 & 3 \\ \cline{2-4} 
& \textbf{Trace2Inv (CVU)} & 2 & 1 \\ \cline{2-4}
& \textbf{Trace2Inv (DFU)} & 1 & 1 \\ \cline{2-4}
& \textbf{Trace2Inv (DFL)} & 1 & 1 \\ \cline{2-4}
& \textbf{InvSol (DDFTI)} & 4 & 4 \\ \cline{2-4} 
& \textbf{InvSol (DFDI)} & 3 & 2 \\ \hline 

\multirow{4}{2.0cm}{\textbf{Gas Control}} 
& \textbf{Trace2Inv (GS)} & 8 & 7 \\ \cline{2-4}
& \textbf{Trace2Inv (GC)} & 7 & 6 \\ \cline{2-4}
& \textbf{InvSol (DGTI)} & 9 & 6 \\ \cline{2-4} 
& \textbf{InvSol (GERI)} & 8 & 7 \\ \hline

\multirow{7}{2.0cm}{\textbf{Access Control}} 
& \textbf{Trace2Inv (EOA)} & 7 & 4 \\ \cline{2-4}
& \textbf{Trace2Inv (SO)} & 6 & 4 \\ \cline{2-4}
& \textbf{Trace2Inv (OO)} & 4 & 3 \\ \cline{2-4}
& \textbf{Trace2Inv (SM)} & 4 & 2 \\ \cline{2-4}
& \textbf{Trace2Inv (OM)} & 2 & 2 \\ \cline{2-4}
& \textbf{InvSol (RBACI)} & 9 & 6 \\ \cline{2-4}
& \textbf{InvSol (MSAI)} & 8 & 6 \\ \cline{2-4}
& \textbf{InvSol (OTVI)} & 7 & 4 \\ \hline 

\multirow{3}{2.0cm}{\textbf{Time Lock}} 
& \textbf{Trace2Inv (SB)} & 5 & 3 \\ \cline{2-4}
& \textbf{Trace2Inv (OB)} & 4 & 2 \\ \cline{2-4}
& \textbf{Trace2Inv (LU)} & 3 & 2 \\ \cline{2-4}
& \textbf{InvSol (BTI)} & 2 & 2 \\ \cline{2-4} 
& \textbf{InvSol (FCFI)} & 2 & 1 \\ \cline{2-4}
& \textbf{InvSol (SBI)} & 2 & 1 \\ \hline

\multirow{3}{2.0cm}{\textbf{Oracle Slippage}} 
& \textbf{Trace2Inv (OR)} & 6 & 4 \\ \cline{2-4}
& \textbf{Trace2Inv (OD)} & 5 & 3 \\ \cline{2-4}
& \textbf{InvSol (DODI)} & 4 & 2 \\ \cline{2-4}
& \textbf{InvSol (PII)} & 4 & 1 \\ \hline

\multirow{4}{2.0cm}{\textbf{Money Flow}} 
& \textbf{Trace2Inv (TIU)} & 7 & 5 \\ \cline{2-4}
& \textbf{Trace2Inv (TOU)} & 6 & 4 \\ \cline{2-4} 
& \textbf{Trace2Inv (TIRU)} & 5 & 3 \\ \cline{2-4}
& \textbf{Trace2Inv (TORU)} & 4 & 2 \\ \cline{2-4}
& \textbf{InvSol (CFTI)} & 3 & 3 \\ \cline{2-4} 
& \textbf{InvSol (RLFI)} & 3 & 2 \\ \cline{2-4}
& \textbf{InvSol (DFLI)} & 3 & 1 \\ \hline

\end{tabular}
\caption{InvSol Vs. Trace2Inv on \textit{TestCrafter} test cases}
\label{tab:invsol_vs_trace2inv_2}
\end{table}

\subsection{Time Analysis}

The time taken for invariant generation by \textsf{InvSol} is influenced by multiple factors, including the inherent complexity and size of the Solidity code. Constructs such as loops, recursion, and complex control structures necessitate extensive computation, increasing the processing time. The type and number of invariants, particularly those involving computation-heavy constructs like mappings, modifiers, and contracts, contribute significantly to the time complexity. Additionally, the thoroughness of test case generation, ensuring comprehensive coverage of execution paths and boundary conditions, further impacts the overall time. 

To rigorously analyze the impact of varying code constructs on the execution time of \textsf{InvSol}, we classified the program into three distinct levels of impact—low (\(l\)), medium (\(m\)), and high (\(h\))—based on their observed time complexity. Our empirical analysis reveals that constructs such as enums, structs, functions, state variables, events, and basic arithmetic operations exhibit minimal influence on the overall execution time, thus they are categorized under the low-impact category. Conversely, medium-impact constructs, including mappings, modifiers, inheritance, libraries, interfaces, and more complex control structures like loops and conditional statements, contribute moderately to the execution time. High-impact constructs, such as contracts, payable functions, fallback/receive functions, storage/memory keywords, Solidity data types involving dynamic arrays and strings, and modifiers involving transaction conditions or external calls, are identified as significantly affecting performance.

Therefore, the total effective weights of Solidity codes are calculated using the following formula:
 
\[
W = \sum_{i=1}^{n} (\textsf{Weight}(S_i^x) + \alpha_i)
\]
where \(n\) represents the total number of constructs, \(S_i^x\) denotes the \(i\)th construct in the impact category \(x\), and \(x\) ranges over \{l, m, h\}. The function \(\textsf{Weight}(S_{x_{i}})\) provides the impact score of \(S_i^x\) within its category, while \(\alpha_{i}\) adds a complexity weight factor that considers the number of variables used and defined. In our experiment, we set the weights as \(\text{Weight}(S_i^l) = 1\), \(\text{Weight}(S_i^m) = 2\), and \(\text{Weight}(S_i^h) = 4\).

\begin{figure}[ht]
\centering
\begin{tikzpicture}[scale=0.75] 
\begin{axis}[
    title style={font=\small},
    label style={font=\small},
    tick label style={font=\small},
    legend style={font=\small},
    title={Time Taken vs. Effective Weight of Code},
    xlabel={Effective Weight of code},
    ylabel={Time Taken (minutes)},
    grid=major,
    scaled ticks=false,
    xmin=100, xmax=3000,
    ymin=4, ymax=6, 
    ytick={4, 4.5, 5, 5.5, 6}, 
    yticklabel style={font=\tiny} 
]
\addplot[only marks, mark=*, fill=blue!50!cyan, opacity=0.5] coordinates {
    (205, 4.16) (200, 4.19) (234, 4.12) (256, 4.57) (290, 4.25)
    (313, 4.22) (432, 3.50) (489, 4.00) (546, 4.37) (582, 4.43)
    (699, 4.39) (703, 4.46) (749, 5.39) (807, 4.57) (850, 4.58)
    (900, 5.14) (967, 4.37) (1010, 4.00) (1065, 4.04) (1111, 4.56)
    (1154, 4.59) (1209, 4.22) (1262, 4.60) (1321, 4.24) (1373, 4.12)
    (1501, 4.71) (1446, 4.69) (1521, 4.75) (1602, 4.72) (1661, 4.78)
    (1807, 4.81) (1873, 4.79) (1800, 4.84) (1888, 5.11) (1869, 4.85)
    (2000, 4.55) (2047, 4.39) (2035, 4.99) (2160, 5.03) (2225, 5.02)
    (2279, 5.08) (2338, 5.07) (2394, 5.10) (2459, 5.15) (2513, 5.12)
    (2576, 5.31) (2682, 5.25) (2700, 4.76) (2758, 5.69) (2858, 5.32)
    (2976, 5.36) (2957, 5.38) (2970, 5.40) (2960, 5.43)
    (2208, 5.11) (2384, 5.11) (2483, 5.32) (2574, 5.26) (2769, 5.31)
    (2784, 5.44) (2968, 5.38) (2957, 5.40)
};
\end{axis}
\end{tikzpicture}
\caption{Invariant generation time Vs effective weight of different smart contracts}
\label{fig:loc_vs_time}
\end{figure}

Figure \ref{fig:loc_vs_time} illustrates the relationship between invariant generation time and the effective weight of Solidity code. As shown, the time required generally increases with the effective weight, ranging from approximately 4 minutes for lower weights to about 5.59 minutes for higher weights. This trend indicates that while the time increases with the complexity of the code, the overall execution time of \textsf{InvSol} remains within a manageable range.

\begin{figure}[h!]
    \centering
    \resizebox{\columnwidth}{!}{ % Resize the figure to fit the column width
        \begin{tikzpicture}
            \pie[color={lighterblue, lighterorange, lighteryellow, lightergray, lightermaroon},
                explode=0.1, % Slight separation for better clarity
                radius=3, % Ensures there's enough room for labels
                text=pin, % Position all labels with pins
                pin distance=1cm, % Distance of pin labels from the edge of the pie
                every pin/.style={align=center, font=\large, text width=4cm}, % Explicitly set larger font size here
                every pin edge/.style={<-, thick}, % Styles for the lines connecting pins
                before number=\phantom, % Prevent numbers from being shown inside the pie
                after number=] % Close any prefix from before number
            {
                8.40/Code Translator + Post Processor (8.40\%),
                2.59/AutoJML (2.59\%),
                58.75/Test Crafter (58.75\%),
                18.69/Invariant Generation Module (18.69\%),
                11.57/Static Code Auditor (11.57\%)
            }
        \end{tikzpicture}
    }
    \caption{Time share of each module in \textsf{InvSol}}
    \label{fig:time_distribution}
\end{figure}

Figure \ref{fig:time_distribution} shows the average percentage of time taken by different modules of \textsf{InvSol}. The \textit{Test Crafter} module takes the most time at approximately 58.75\%, reflecting the extensive testing performed to ensure comprehensive coverage. It is  followed by \textit{Invariant Generation Module} module at 18.69\%, \textit{Static Code Auditor} module at 11.57\%. The \textit{Code Translator + Post Processor} and \textit{AutoJML} modules take the least time at 8.40\% and 2.59\% respectively.

\section{Related Work}
\label{label:related-works}

\emph{Daikon} \cite{daikon_tool} is an open-source tool that allows dynamic detection of likely program invariants, offering an implementation that has been widely used for generating program specifications and formal verification across various programming languages. \emph{DIDUCE} \cite{hangal2002tracking} is another invariant detection tool for Java that dynamically checks program behavior for violations of hypothesized invariants. In Solidity, \emph{InvCon} \cite{liu2022invcon} is a dynamic invariant detection tool for Ethereum smart contracts, designed to identify and verify contract invariants from historical transaction data. \emph{InvCon+} \cite{liu2024automated} extends its predecessor \emph{InvCon} by combining dynamic inference and static verification to produce verified and expressive contract invariants. \emph{Cider} \cite{liu2022learning} is a reinforcement learning-based tool designed to automatically learn contract invariants for Solidity smart contracts, with a particular focus on proving arithmetic safety by predicting likely invariants that aid in verification. \emph{Trace2Inv} \cite{chen2024demystifying}  presents a comprehensive study on the effectiveness of runtime invariants for securing smart contracts based on the transaction history, outperforming existing methods in blocking exploits with minimal gas overhead.

\section{Conclusion}
\label{label:conclusion}

In this paper, we presented \textsf{InvSol}, a framework for invariant generation in Solidity smart contracts prior to their deployment. \textsf{InvSol} demonstrated high precision and recall for varied invariant types, compared to the existing tools \emph{InvCon} and \emph{InvCon+}, especially in detecting loop invariants or handling newly deployed contracts. Further, it shows a notable performance improvement in case of vulnerable smart contracts. In future, we plan to extend our proposal to other smart contract languages as well. 

\section*{Acknowledgement}
This research is partially supported by the Core Research Grant (CRG/2022/005794) from the Science and Engineering Research Board (SERB), Department of Science and Technology, Government of India.

\bibliography{references}

\begin{thebibliography}{10}

\bibitem{chen2024demystifying}
Zhiyang Chen et~al.
\newblock Demystifying invariant effectiveness for securing smart contracts.
\newblock {\em arXiv:2404.14580}, 2024.

\bibitem{swc_registry}
ConsenSys Diligence.
\newblock Swc registry: Smart contract weakness classification, 2024.
\newblock Accessed: 2024-09-02.

\bibitem{evosuite}
{EvoSuite Developers}.
\newblock Evosuite documentation, 2024.
\newblock Accessed: 2024-07-20.

\bibitem{ethereum_eip}
Ethereum Foundation.
\newblock Ethereum improvement proposals (eip), 2024.
\newblock Accessed: 2024-09-02.

\bibitem{galeotti2014dynamate}
Juan~P. Galeotti, Carlo~A. Furia, Eva May, Gordon Fraser, and Andreas Zeller.
\newblock Dynamate: Dynamically inferring loop invariants for automatic full functional verification.
\newblock In {\em HVC 2014: 10th International Haifa Verification Conference}, pages 48--53. Springer, 2014.

\bibitem{google_bigquery}
{Google Cloud}.
\newblock Google bigquery.
\newblock \url{https://cloud.google.com/bigquery}.

\bibitem{hangal2002tracking}
Sudheendra Hangal and Monica~S Lam.
\newblock Tracking down software bugs using automatic anomaly detection.
\newblock In {\em Proceedings of the 24th international conference on Software engineering}, pages 291--301, 2002.

\bibitem{hoare1969axiomatic}
Charles Antony~Richard Hoare.
\newblock An axiomatic basis for computer programming.
\newblock {\em Communications of the ACM}, 12(10):576--580, 1969.

\bibitem{jml_language}
{JML Project}.
\newblock Java modeling language (jml).
\newblock \url{https://www.eecs.ucf.edu/~leavens/JML/}.

\bibitem{liu2022learning}
Junrui Liu, Yanju Chen, Bryan Tan, et~al.
\newblock Learning contract invariants using reinforcement learning.
\newblock In {\em 37th IEEE/ACM ASE}, pages 1--11, 2022.

\bibitem{liu2022invcon}
Ye~Liu and Yi~Li.
\newblock Invcon: A dynamic invariant detector for ethereum smart contracts.
\newblock In {\em 37th IEEE/ACM ASE}, pages 1--4, 2022.

\bibitem{liu2024automated}
Ye~Liu, Chengxuan Zhang, et~al.
\newblock Automated invariant generation for solidity smart contracts.
\newblock {\em arXiv:2401.00650}, 2024.

\bibitem{nikolic2018finding}
Ivica Nikoli{\'c}, Aashish Kolluri, Ilya Sergey, Prateek Saxena, and Aquinas Hobor.
\newblock Finding the greedy, prodigal, and suicidal contracts at scale.
\newblock In {\em Proceedings of the 34th annual computer security applications conference}, pages 653--663, 2018.

\bibitem{openzeppelin_contracts}
OpenZeppelin.
\newblock Openzeppelin contracts, 2024.
\newblock Accessed: 2024-09-02.

\bibitem{daikon_tool}
Jeffrey~H. Perkins, Michael~D. Ernst, et~al.
\newblock Daikon invariant detector.
\newblock \url{https://plse.cs.washington.edu/daikon/}.

\bibitem{smartbugs_dataset}
Smartbugs Project.
\newblock Smartbugs: A curated dataset of solidity smart contracts.
\newblock \url{https://smartbugs.github.io}.

\bibitem{so2020verismart}
Sunbeom So, Myungho Lee, Jisu Park, Heejo Lee, and Hakjoo Oh.
\newblock Verismart: A highly precise safety verifier for ethereum smart contracts.
\newblock In {\em 2020 IEEE Symposium on Security and Privacy (SP)}, pages 1678--1694. IEEE, 2020.

\end{thebibliography}

\end{document}